\documentclass[apj]{emulateapj}
\usepackage{lineno}
\usepackage{epsfig}
\usepackage{graphicx}
\usepackage[utf8]{inputenc}
\usepackage[T1]{fontenc}
\usepackage[english]{babel}
\usepackage{booktabs}
\usepackage{epstopdf}
\usepackage{amsmath}
\usepackage{natbib}
\usepackage{hyperref}
\usepackage{lipsum}
\usepackage{enumerate}
\usepackage{figsize}
\shorttitle{Realistic AGN Mock Catalogs}
\shortauthors{Allevato et al.}

\begin{document}

\title{Building robust AGN mock catalogs to unveil black hole evolution and for survey planning}

\author{V. Allevato\altaffilmark{1,2,3},
F. Shankar\altaffilmark{4},
C. Marsden \altaffilmark{4},
U. Rasulov \altaffilmark{4},
A. Viitanen \altaffilmark{3,6},
A. Georgakakis \altaffilmark{5},
A. Ferrara \altaffilmark{1},
A. Finoguenov \altaffilmark{3}}

\altaffiltext{1}{Scuola Normale Superiore, Piazza dei Cavalieri 7, I-56126 Pisa, Italy}
\altaffiltext{2}{INAF - Osservatorio di Astrofisica e Scienza delle Spazio di Bologna, OAS, Via Gobetti 93/3, 40129, Bologna, Italy}
\altaffiltext{3}{Department of Physics, University of Helsinki, PO Box 64, FI-00014 Helsinki, Finland}
\altaffiltext{4}{Department of Physics and Astronomy, University of Southampton, Highfield SO17 1BJ, UK}
\altaffiltext{5}{Institute for Astronomy \& Astrophysics, National Observatory of Athens, V. Paulou \& I. Metaxa, 11532, Greece}
\altaffiltext{6}{Helsinki Institute of Physics, Gustaf Hällströmin katu 2, University of Helsinki, Finland}

\begin{abstract}

The statistical distributions of active galactic nuclei (AGN), i.e.\ accreting supermassive black holes (BHs), in mass, space and time, are controlled by a series of key properties, namely the BH-galaxy scaling relations, Eddington ratio distributions and fraction of active BH (duty cycle). Shedding light on these properties yields strong constraints on the AGN triggering mechanisms whilst providing a clear baseline to create useful ``mock'' catalogues for the planning of large galaxy surveys. We here delineate a robust methodology to create mock AGN catalogs built on top of large N-body dark matter simulations via state-of-the-art semi-empirical models. We show that by using as independent tests the AGN clustering at fixed X-ray luminosity, galaxy stellar mass and BH mass, along with the fraction of AGN in groups and clusters, it is possible to significantly narrow down the choice in the relation between black hole mass and host galaxy stellar mass, the duty cycle, and the average Eddington ratio distribution, delivering well-suited constraints to guide cosmological models for the co-evolution of BHs and galaxies. Avoiding such a step-by-step methodology inevitably leads to strong degeneracies in the final mock catalogs, severely limiting their usefulness in understanding AGN evolution and in survey planning and testing.
\end{abstract}


\section{Introduction}
\label{sec:intro}

Several semi-analytical models and hydrodynamical simulations
\citep[e.g.][]{springel05,hopkins06,menci08}
have been developed in recent years to describe the main mechanisms
that fuel the central supermassive black holes (BHs). With suitable adjustment of parameters,
these models can explain many aspects of AGN phenomenology
\citep[e.g.][]{hopkins06,hopkins08}.
Often relying on a rather heavy parameterization of the physics regulating the cooling, star formation, feedback, and merging of baryons
\citep[e.g.][]{monaco07},
semi-analytic models of galaxy evolution can present serious degeneracies
\citep[e.g.][]{gonzalez11,lapi18},
or even significant divergences in, e.g., the adopted sub-grid physics
\citep[][]{scannapieco12,nunez-castineyra20}.
Semi-empirical models (SEMs) represent an original
and complementary methodology to more traditional modelling approaches
\citep[e.g.][]{hopkins09b}.
The aim of SEMs is to tackle specific aspects of galaxy and BH evolution in a transparent, fast, and flexible way, relying on just a few input assumptions and parameters. SEMs cannot replace ab-initio models of galaxy and BH evolution but can provide guidance to reduce the space of parameters and shed light on the viable physical processes.

\begin{figure*}
	\plottwo{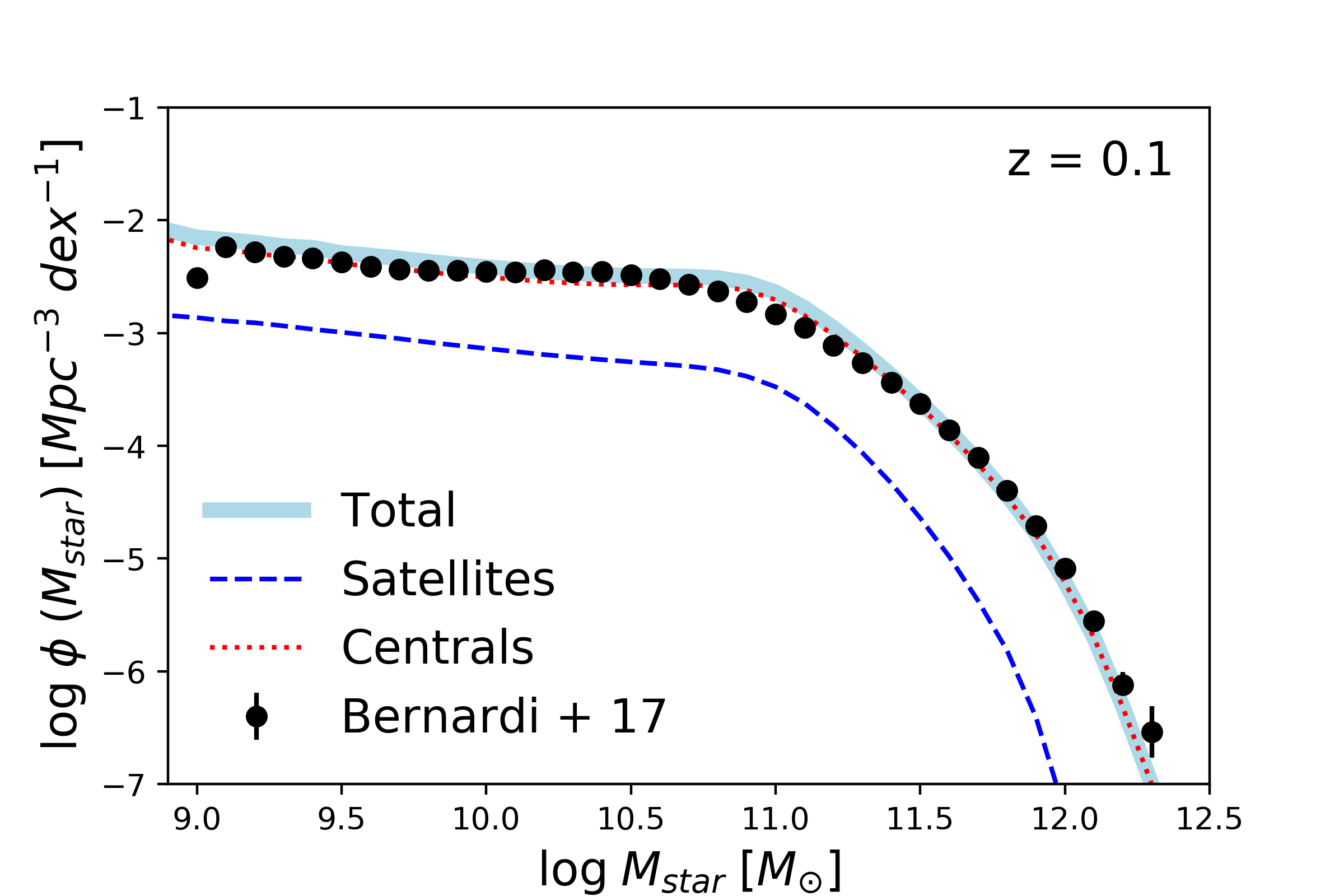}{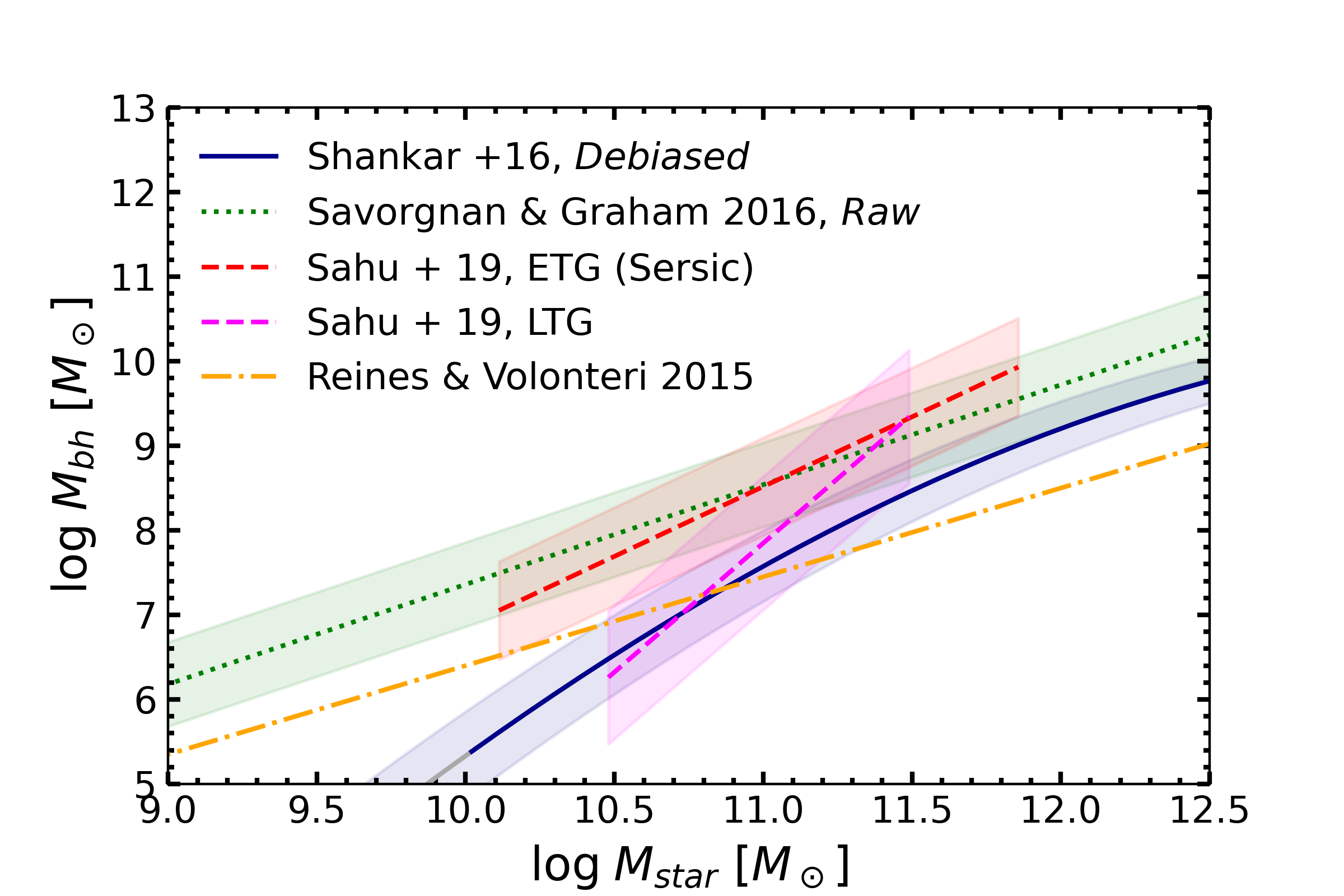}
	\caption{\footnotesize Left Panel: Stellar mass function at z = 0.1 for central and satellite mock galaxies, compared with measurements by using SDSS-DR7 galaxies. Right Panel: BH mass--stellar mass relation, as put forward by
    \citet[][\emph{debiased} ]{shankar16}
    and as derived for local galaxy samples with dynamically measured BH masses from
    \citet[][\emph{raw}]{savorgnan16}.
  The M$_{\mathrm{bh}}-M_{\mathrm{star}}$ relations as derived for early type and late type galaxy by
    \citet[][]{sahu19}, and for AGN by \citet{reines15}
are shown for comparison.}
\label{fig1}
\end{figure*}

It is particularly relevant the application of SEMs to the creation of active and normal galaxy ``mock'' catalogues \citep[e.g.][]{conroy13},
which are a vital component of the planning of imminent
extra-galactic surveys such as Euclid \citep[][]{laureijs11}
and the Vera C. Rubin Observatory Legacy Survey of Space and Time
\citep[LSST;][]{lsst09}.
The first step for the creation of mocks consists in
assigning galaxies to dark matter halos extracted
from large cosmological N-body simulations
\citep[e.g.][]{riebe13,klypin16},
via abundance-matching techniques
\citep[e.g.][]{kravtsov04,vale04,shankar06,behroozi13a,moster13}.
Despite being based on minimal assumptions, the latter are not
immune to important systematics, mostly related to the input
data, which propagate onto the star formation and mass
assembly histories predicted by SEMs
\citep[e.g.][]{grylls20a,grylls20b,oleary20}.

In the last few years, several studies have focused on the creation of mock catalogs specifically for AGN that can be utilized for the planning and testing of large-scale AGN-dedicated extragalactic surveys such eROSITA
\citep[e.g.][]{georgakakis19,comparat19,aird20}.
These AGN mocks are built by starting from an empirical galaxy catalog and by
assigning to each object a specific accretion-rate that is proportional to the quantity $L_X/M_{\mathrm{star}}$, drawn randomly from observationally determined probability distributions $P_\mathrm{AGN}(L_X/M_{\mathrm{star}})$
\citep[e.g.][]{bongiorno16,georgakakis17,aird18}.
This quantity can be measured directly from observations and provides an estimate of X-ray emission per unit stellar mass for a galaxy.
\begin{figure}
	\plotone{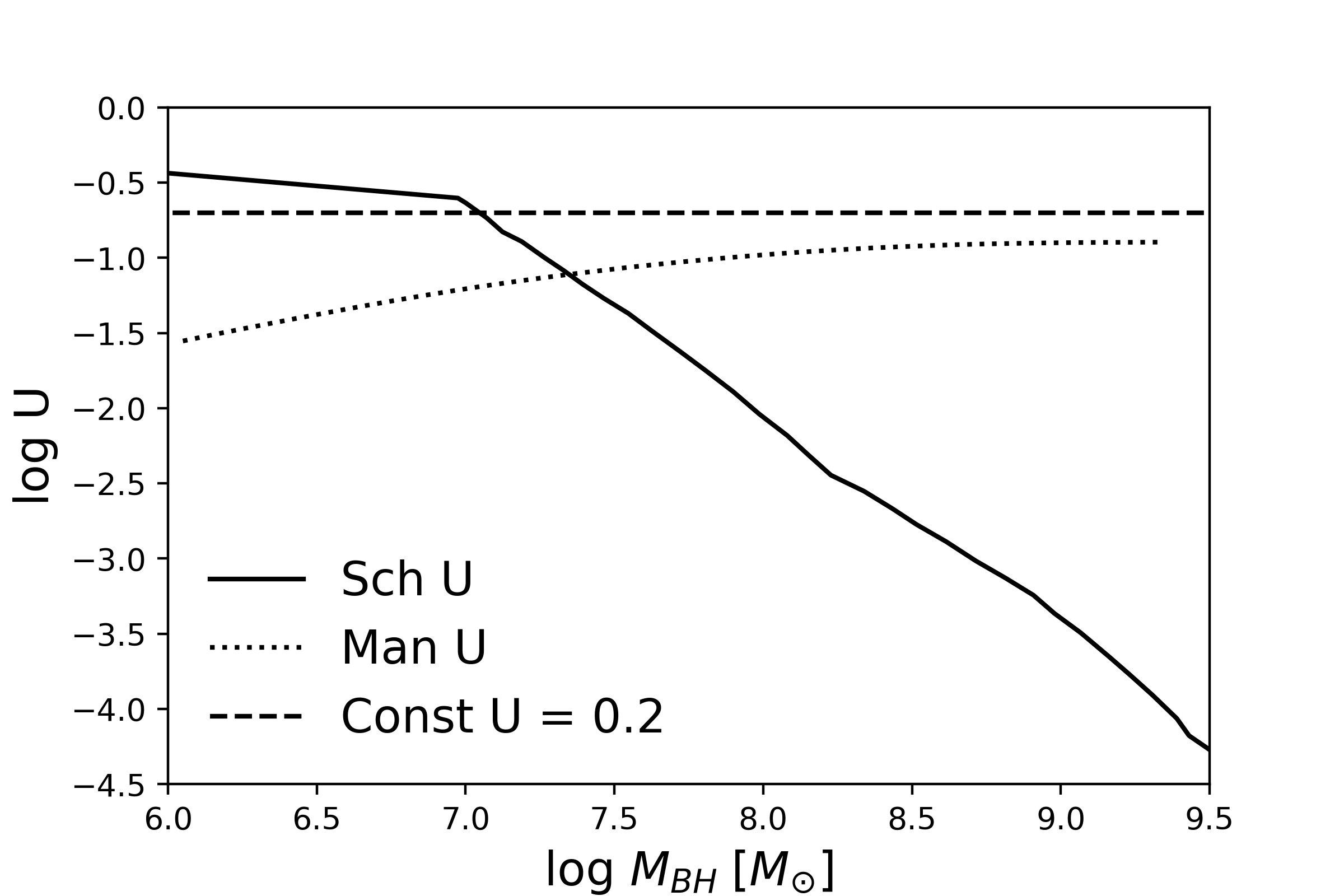}
	\caption{\footnotesize Duty cycle U as a function of M$_\mathrm{BH}$ as derived in
  \citet[][continuous line]{schulze10}
  at z = 0.1,
  \citet[][dotted line]{man19}
  at z $<$ 0.1 and
  \citet[][dashed line]{goulding10}.}
	\label{fig3}
\end{figure}
The advantage of this methodology is that by using
just a few input relations, namely the stellar mass-halo
mass relation (from abundance matching) and the probability
distribution of specific accretion-rate $P_\mathrm{AGN}$,
it allows to create a mock catalog of AGN that
-- by design -- reproduces the observed X-ray luminosity
function (XLF) and broadly matches the large-scale bias at a given host
galaxy stellar mass. 
Using this approach, \citet{georgakakis19} 
populated cosmological simulations with AGN and
showed that their clustering properties (including the signal at small scales), are consistent with
state-of-the-art observational measurements of X-ray or
UV/optically selected samples at different redshifts and accretion luminosities, supporting the view that the large-scale distribution
of AGN may be independent of the detailed physics of BH fueling. 
Some recent works also tested the same methodology against the large-scale bias dependence 
on the X-ray luminosity at different redshifts \citep{georgakakis19,aird20}. 

However, in these models key information such as BH mass is largely bypassed, and the AGN duty
cycle (i.e.\ the probability for a galaxy of being active
above a certain luminosity or threshold) is not considered as a separate model input parameter, limiting the efficacy of these models
in shedding light on the processes controlling the
co-evolution of BHs and their hosts. Moreover, in
these models the assignment of specific accretion-rates
to mock galaxies by using $P_\mathrm{AGN}$ is a
stochastic process, assumed to be independent of the environment (centrals and satellites of similar stellar mass share the same probability of being active).

In this paper instead, we create mock catalogs of AGN
by varying different input model parameters, namely the
stellar mass--halo mass and BH mass--stellar mass relations, the AGN duty cycle, the Eddington ratio distribution
and the fraction of satellite AGN (controlled by the parameter $Q$ defined later)
and test the effect on several observables, such as the AGN XLF, the $P_\mathrm{AGN}$ distribution and AGN large-scale bias as a function of BH/stellar mass and luminosity. 
More generally, we demonstrate in this study that
by calibrating the AGN mocks on the bias
at fixed BH mass, stellar mass, and AGN luminosity,
provides a self-consistent and robust route
to break the most relevant degeneracies and narrow
down the choice of input parameters. For example,
\citet[][]{shankar20}
emphasized that current measurements of AGN clustering at $z=0.25$
\citep[][]{krumpe15}
are already sufficient to constrain, in ways
independent of the AGN duty cycle, the scaling relations of BHs
\citep[e.g.][]{kormendy13,reines15,savorgnan16,shankar16,davis18}.
The main goal of this paper is to provide a complete framework to build a robust and realistic AGN mock catalog, consistent with many different and independent observables, and physically sound, being
based on the underlying scaling relations between BHs and their host galaxies and dark matter halos.

\begin{figure*}
\centering
\plottwo{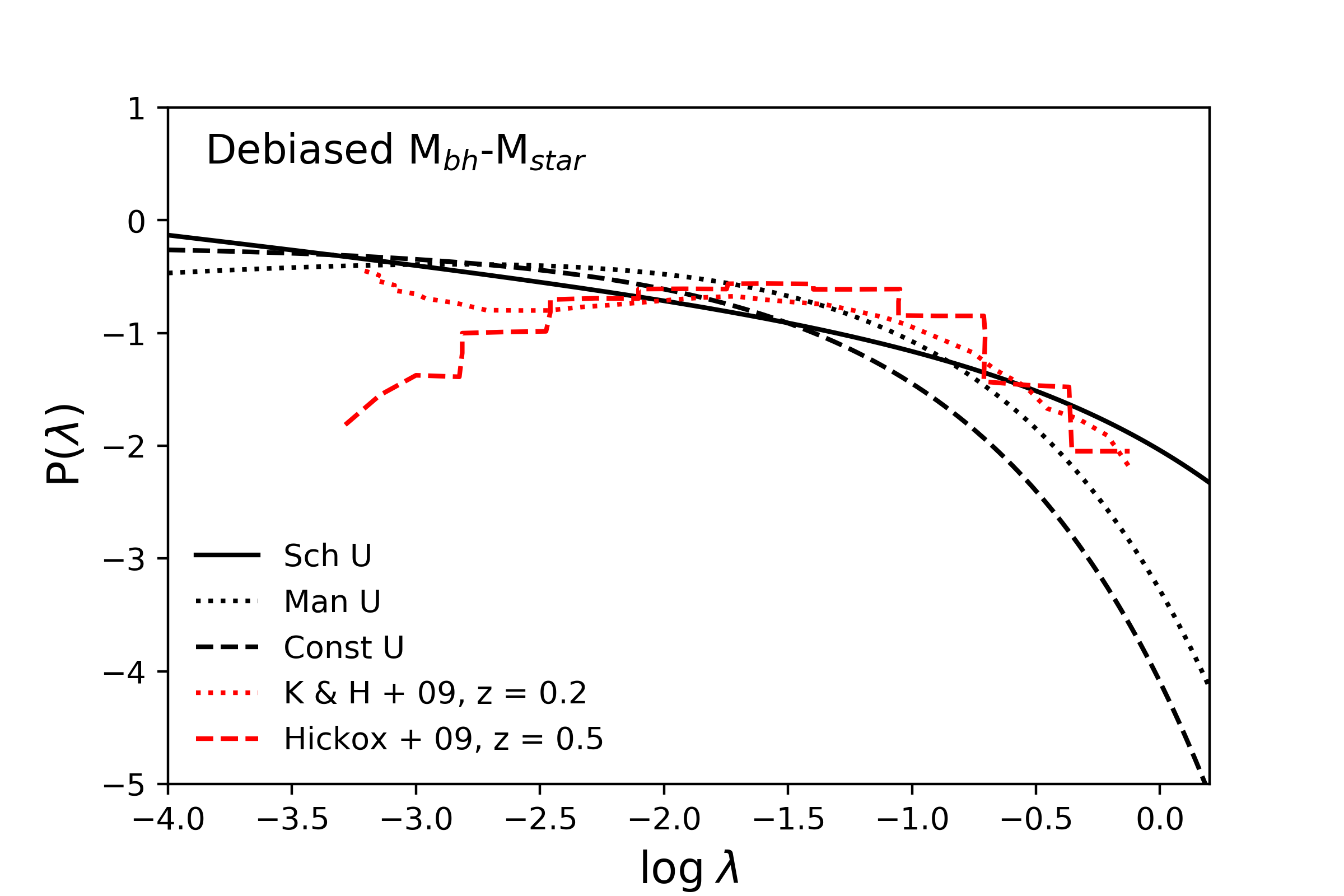}%
        {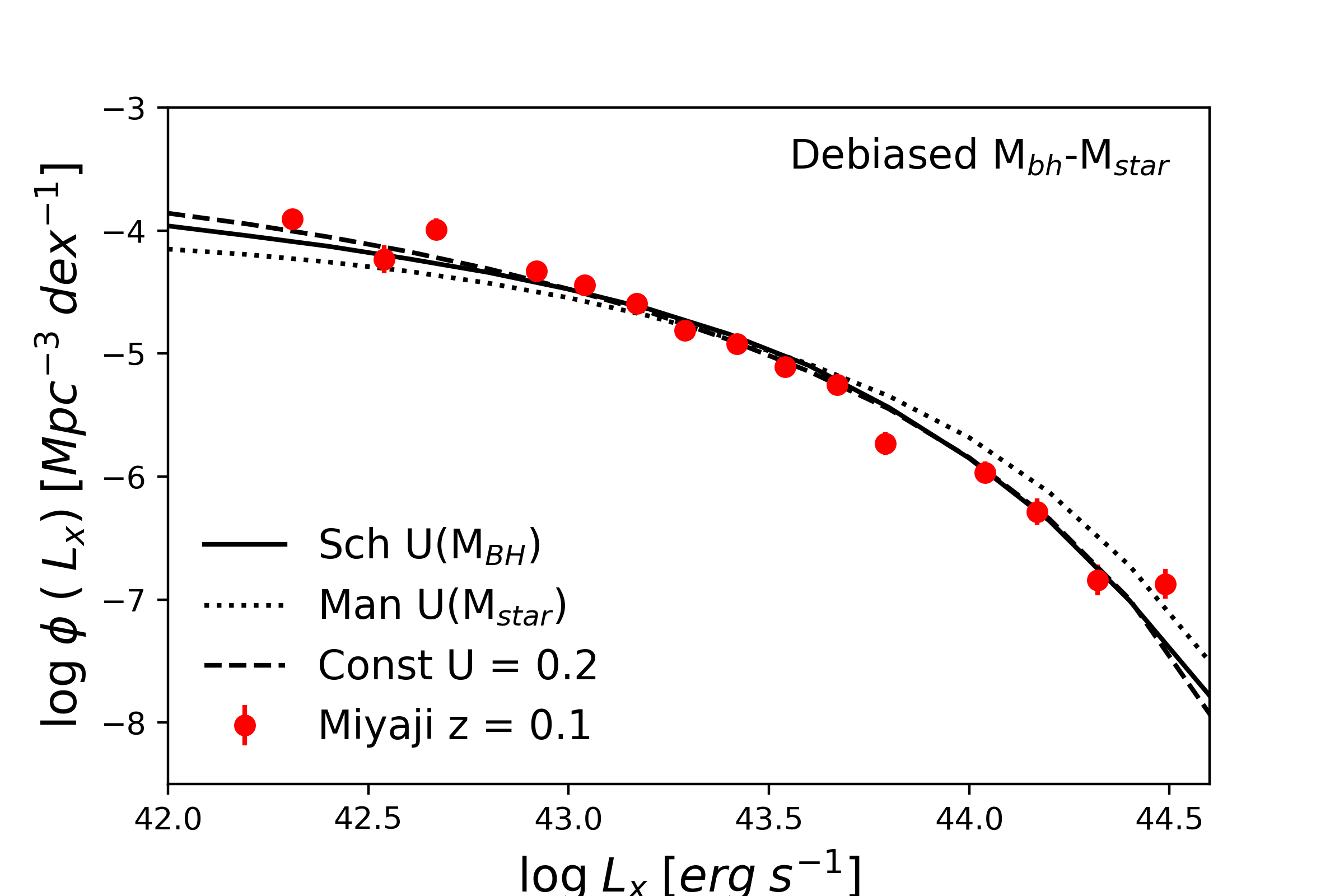}%
\hspace{0.1em}
\caption{\footnotesize Left Panel: Input Eddington ratio distribution $P(\lambda)$, described by a Schechter function characterised by a knee $\lambda^{\star}$ and a power-law index $\alpha$ derived to reproduce the AGN luminosity, when using the BH mass--stellar mass relation from \citet[][debiased]{shankar06} and the AGN duty cycle derived by
\citet[][continuous line]{schulze10},
\citet[][dotted line]{man19},
and \citet[][dashed line]{goulding10},
compared to the results of
\citet[][dotted red line]{Kauffmann09} and \citet[][dotted red line]{hickox09}.
Right panel: Corresponding X-ray luminosity function for mock AGN at z = 0.1 compared to the X-ray luminosity function as derived for Compton Thin un/obscured AGN in
\citet[][]{miyaji15}.
}
\label{fig4}
\end{figure*}

\section{Research Methodology}

In this Section we provide the step-by-step description of our baseline methodology:
\begin{itemize}
\item At a given redshift of reference, in this work $z = 0.1$, we extract large catalogs of DM halos and
subhalos from large, N-body dark matter simulations. We here rely on the MultiDark simulation
\citep[][]{riebe13}.
The catalogues contain both central/parent halos and satellite halos with unstripped mass at infall.
\item To each parent halo a central galaxy is assigned with stellar mass given by abundance matching relations at the redshift of reference
\citep[e.g.][]{grylls19}, while satellite halos are assigned a stellar mass at their redshift of infall.
\item To each galaxy we assign a BH mass from an empirical BH mass-galaxy mass relation drawn from several recent studies
\citep[e.g.][]{shankar16}.
\item To each galaxy and BH we then assign an Eddington ratio $\lambda = L_{bol}/L_{\rm Edd}$, with $L_{bol}$ the bolometric luminosity and $L_{\rm Edd}$ the Eddington limit of the BH\@. The parameter $\lambda$ is randomly extracted from a $P(\lambda)$ distribution described by a Schechter function, the latter chosen in a way to reproduce the AGN XLF at $z = 0.1$, for a given input ``duty cycle'' (see below).
In our reference model we ignore, for simplicity, any mass dependence of $P(\lambda)$ on, e.g., BH mass. We will discuss in Section~\ref{sec6} the (moderate) impact of relaxing this assumption.
Regardless, we note that any mass dependence in $P(\lambda)$ is degenerate with the duty cycle
\citep[e.g.\ in the AGN XLF;][]{shankar13},
a model input parameter which we explore thoroughly in this work.
\item To each galaxy/BH an extinction corrected X-ray luminosity $L_X$ in the 2-10 keV band is then assigned from the bolometric luminosity $L_{bol}$ via up-to-date bolometric corrections
\citep[e.g.][]{duras20}.
\item Each galaxy and its associated BH is assigned a duty cycle, i.e.\ a probability for a BH of a given M$_\mathrm{BH}$ of being active, following empirically-based duty cycles
\citep[e.g.][]{man19}.
\end{itemize}

We then generate our mock catalog of AGN and, by varying our input parameters, test a number of outputs, such as the AGN XLF, the AGN specific accretion-rate distribution $P_\mathrm{AGN}$ and the AGN large-scale clustering. We focus on $z = 0.1$ where the galaxy-BH scaling relations are better constrained and additional measurements on some of the key observables, such as AGN-galaxy clustering, are available. We stress that the methodology we put forward in this work is applicable to any redshift of interest. In Viitanen et al. (submitted), for example, we apply our methodology to $z \sim 1.2$, while in Carraro et al. (submitted) we push our methodology up to $z\sim 3$, and specifically focus on the correlation with star formation rate, which is not explicitly included in the present work.

\begin{figure*}
\centering
\plottwo{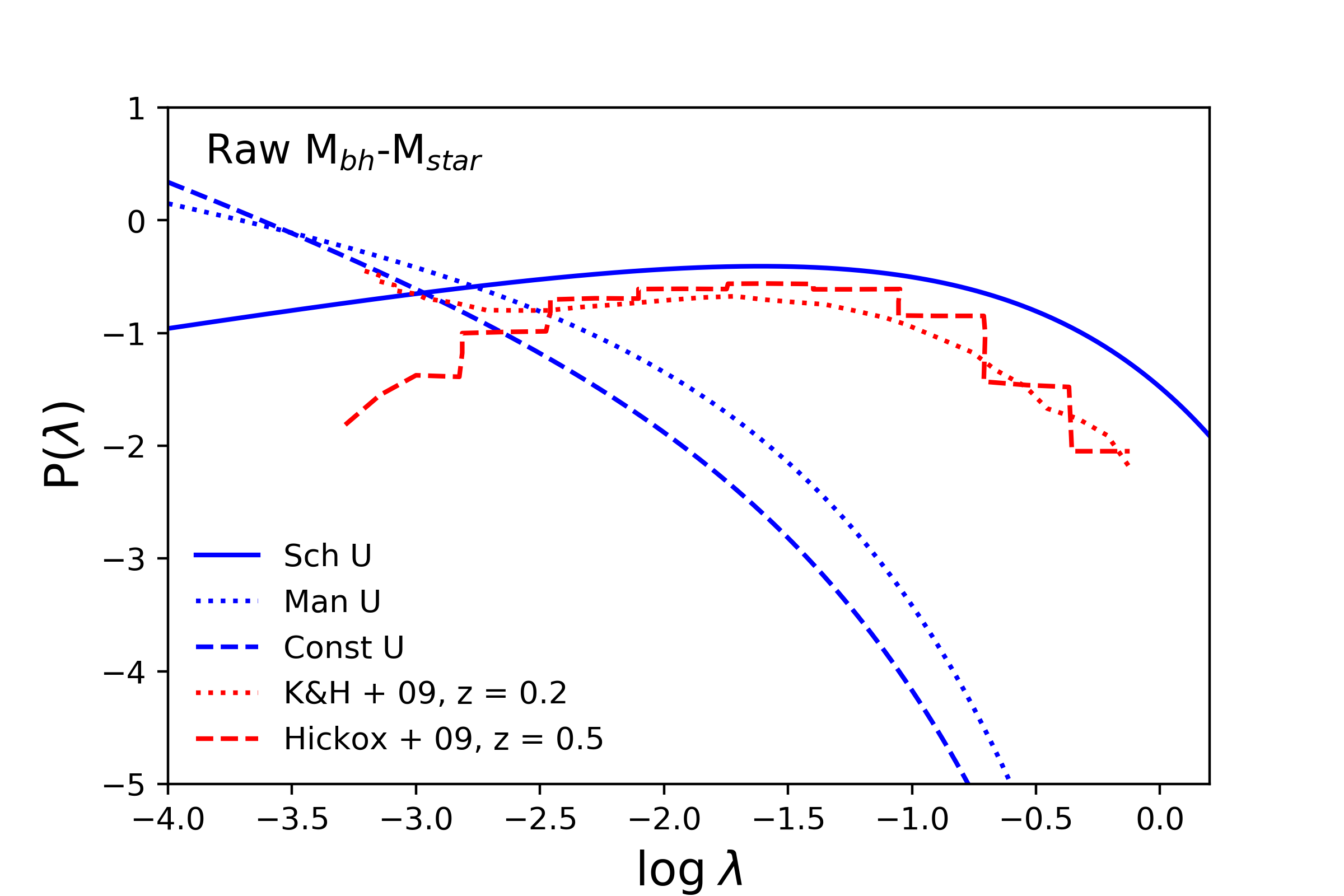}%
        {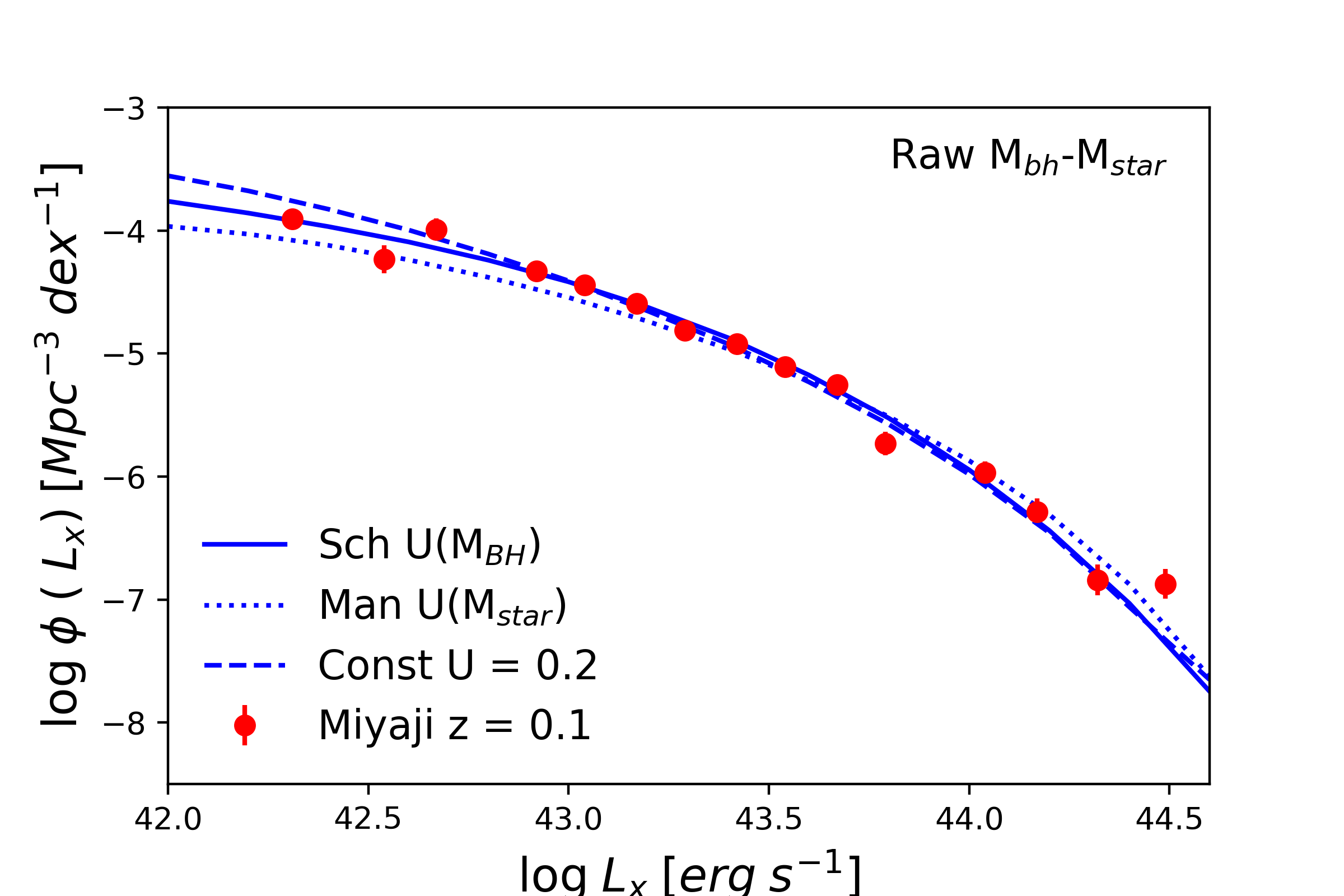}%
\caption{\footnotesize Left Panel: Input Eddington ratio distribution $P(\lambda)$, described by a Schechter function characterised by a knee $\lambda^{\star}$ and a power-law index $\alpha$ derived to reproduce the AGN luminosity function, when using the BH mass--stellar mass relation from \citet[][raw]{savorgnan16} and the AGN duty cycle derived by
\citet[continuous line,][]{schulze10},
\citet[dotted line,][]{man19}
and assuming a constant $U = 0.20$
\citep[][dashed line]{goulding10}.
compared to the results of
\citet[][dotted red line]{Kauffmann09} and \citet[][dashed red line]{hickox09}.
Right panel: Corresponding X-ray luminosity function for mock AGN at z = 0.1 compared to the X-ray luminosity function as derived for Compton Thin un/obscured AGN in
\citet[][]{miyaji15}.
}
\label{fig5}
\end{figure*}

\subsection{Connecting halos to galaxies and BHs}

We start from a large catalog of dark matter halos and
subhalos from MultiDark\footnote{www.cosmosim.org}-Planck 2
\citep[MDPL2;][]{riebe13}
at the redshift z = 0.1.
MDPL2 currently provides the largest publicly available
set of high-resolution and large volume N-body simulations (box size of 1000 h$^{-1}$Mpc, mass resolution of 1.51$\times$10$^9$ h$^{-1}$M$_{\odot}$).
The \texttt{ROCKSTAR} halo finder
\citep[][]{behroozi13b}
has been
applied to the MDPL2 simulations to identify halos and flag those
(sub-halos) that lie within the virial radius of a more massive host halo.
The mass of the dark matter halo is defined as the virial mass in the case of
host halos and the infall progenitor virial mass for sub-halos.

From abundance matching techniques one can infer the
stellar mass--halo mass relation which shows that the baryons
are converted into stars with very different efficiencies
in halos of diverse mass
\citep[e.g.][]{shankar06,moster13}.
We adopt the parameterization for the stellar-to-halo mass ratio by
\citet[][]{moster13}:
\begin{equation}\label{eq1}
M_{\mathrm{star}} (M_h, z) = 2 M_h N {\left[ \frac{M_h}{M_n(z)}^{-\beta(z)} + \frac{M_h}{M_n(z)}^{\gamma(z)} \right]}^{-1}
\end{equation}
where $N$ is the normalization of the stellar-to-halo mass ratio,
$M_n$ a characteristic mass where the ratio is equal to
the normalization $N$, and two slopes $\beta$ and $\gamma$
which indicate the behavior at low and high-halo mass ends, respectively. We fixed these redshift-dependent parameters
as in \citet[][]{grylls19}
who suggested a steeper slope than
\citet[][]{moster13} for the high-mass end
\citep[as also shown in][]{shankar14,shankar17,kravtsov18},
which better fits the SDSS-DR7 from \citet[][]{meert15,meert16},
with improved galaxy photometry
\footnote{We decrease by 0.1 dex the original stellar masses by
\citet[][]{grylls19}
to further improve the match to the latest stellar mass function by
\citet[][]{bernardi17}.}.

Figure~\ref{fig1} (left panel) shows the stellar
mass functions at $z=0.1$ presented in
\citet[][]{bernardi17},
based on S\'{e}rsic-exponential
fits to the surface brightness profiles of
galaxies in the Sloan Digital Sky Survey (SDSS) Data Release 7 (DR7), 
and characterized by a significantly higher number densities of massive galaxies ($>10^{12}$ M$_{\ast}$)
when compared to estimates by, e.g.,
\citet[][]{baldry12,moustakas13,bernardi10,bell03}.

\begin{figure*}
\centering
\plottwo{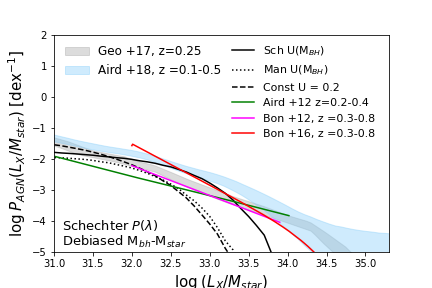}%
        {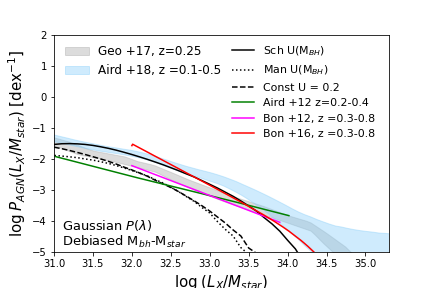}%
\caption{Specific accretion-rate distribution $P_\mathrm{AGN} (\lambda \propto L_X/M_{\mathrm{star}})$ defined as the probability that a galaxy of a given $\lambda$ is an AGN, and given by the convolution of the input Eddington ratio distribution $P(\lambda)$ and the AGN duty cycle (Eq.~\ref{eq9}). The prediction from mock AGNs assuming a Schechter (left panel) and Gaussian (right panel) input $P(\lambda)$ and a $M_{\mathrm{bh}}-M_{\mathrm{star}}$ relation as defined in \citet[][debiased]{shankar06} are compared with data from
\citet[][]{aird12,aird18,bongiorno12,bongiorno16,georgakakis17},
according to the legend.
}
\label{fig6}
\end{figure*}
\begin{figure*}
\centering
\plottwo{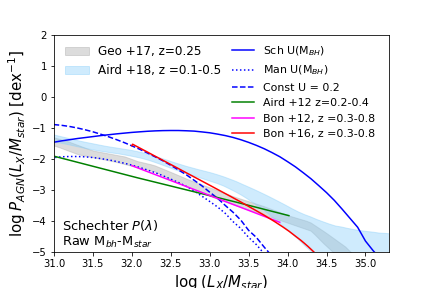}%
        {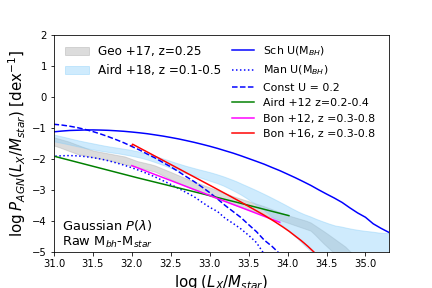}%
\caption{Specific accretion-rate distribution $P_\mathrm{AGN} (\lambda \propto L_X/M_{\mathrm{star}})$ defined as the probability that a galaxy of a given $\lambda$ is an AGN, and given by the convolution of the input Eddington ratio distribution $P(\lambda)$ and the AGN duty cycle (Eq.~\ref{eq9}). The prediction from mock AGNs with a Schechter (left panel) and Gaussian (right panel) input $P(\lambda)$ and a $M_{\mathrm{bh}}-M_{\mathrm{star}}$ relation as defined in \citet[][raw]{savorgnan16} are compared with data from
\citet[][]{aird12,aird18,bongiorno12,bongiorno16,georgakakis17},
according to the legend.}
\label{fig7}
\end{figure*}

\subsection{Input \texorpdfstring{M$_{\mathrm{BH}}$-M$_{\mathrm{star}}$}{mbh-mstar} relation}\label{sec:4}

As a second step, to each galaxy we assign a BH mass assuming the following two scaling relations:
\begin{itemize}
	\item The BH mass--stellar mass relation as derived in
    \citet[][]{shankar16}, labelled as BHSMR-Sha16 ``debiased'' hereafter:
	\begin{eqnarray}
	\log\frac{M_{\mathrm{BH}}}{M_{\odot}} = 7.574 + 1.946 \, \log\left( \frac{M_{\mathrm{star}}}{10^{11}M_{\odot}}\right)  -0.306 \nonumber\\
	\times {\left[ \log \left( \frac{M_{\mathrm{star}}}{10^{11}M_{\odot}}\right) \right]}^2
	-0.011  {\left[ \log \left( \frac{M_{\mathrm{star}}}{10^{11}M_{\odot}}\right) \right]}^3,
	\label{eq2}
	\end{eqnarray}

with a mass-dependent intrinsic scatter given by:
\begin{equation}
\Delta \log \frac{M_{\mathrm{BH}}}{M_{\odot}} = 0.32 - 0.1 \times \log \left( \frac{M_{\mathrm{BH}}}{M_{\odot}} \right)
\label{eq3}
\end{equation}
as presented in Equation 5 of
\citet[][]{shankar19}.
Note that this equation is applicable to galaxies with stellar masses log M$_{\mathrm{star}}/M_{\odot}\ga10$;

\item the relation derived for the
  \citet[][]{savorgnan16}
  sample of galaxies with dynamically measured
  BH masses, BHSMR-SG16 (raw) hereafter, as presented in Equation 3 of
  \citet[][]{shankar19},
  with a scatter of 0.5 dex:
\begin{equation}
\log\frac{M_{\mathrm{BH}}}{M_{\odot}} = 8.54 + 1.18 \, \log \left( \frac{M_{\mathrm{star}}}{10^{11}M_{\odot}}\right) \, .
\label{eq4}
\end{equation}

\end{itemize}

The right panel of Figure~\ref{fig1} shows the M$_{\mathrm{BH}}$-M$_{\mathrm{star}}$ relations defined by Equations~\ref{eq2} and~\ref{eq4} with their associated dispersions and compared with several other relations from the recent literature, as labelled.
It can be seen from Figure~\ref{fig1} that our two chosen relations bracket the systematic uncertainties in both slope and normalization present in the local BH mass-galaxy stellar mass relation.

In our reference models throughout we include the scatters in the relations described above as random normal dispersions. However, it may be possible that some correlation between, in particular, the dispersions in BH mass and galaxy stellar mass at a given DM halo mass may exist. We thus explore in sec~\ref{sec6} some of the main consequences on our results of including a degree of covariance in the scatters, and refer to Viitanen et al. (submitted) for a more comprehensive discussion of implementing a covariant scatter in the input stellar mass--halo mass and BH mass--stellar mass relations.

\subsection{Input Eddington ratio distribution}

To each galaxy and BH we assign an Eddington ratio $\lambda \equiv L_{bol}/L_{Edd}$ following a $P(\lambda$) distribution described by:

\begin{itemize}
    \item a Schechter function:
    \begin{equation}
P(\lambda) \propto {\left( \frac{\lambda}{\lambda^{\star}} \right)}^{-\alpha}  \exp \left( { {-\frac{\lambda}{\lambda^{\star}}}}\right)
\end{equation} \label{eq7sch}
with $\lambda$ in the range $\lambda$ = 10$^{-4}$ -- $10^{1}$.
The Schechter function is characterized by two free parameters: the knee  $\lambda^{\star}$, where the power-law form of the function cuts off and
the power-law index $\alpha$.
The Schechter function is supported by recent studies on the
specific accretion-rate distribution of AGN, such as
\citet[][]{bongiorno16,aird17,aird18,aird19,georgakakis17}.
    \item a Gaussian function:
\begin{equation}
P(\log(\lambda)) \propto \exp{\left( -\frac{{[\log(\lambda) - \mu]}^2 }{2 \sigma^2} \right)}
\end{equation}\label{eq7bis}
where $\log(\lambda)$ varies in the range log$\lambda$ = - 4 -- 1, $\sigma$ is the standard deviation and $\mu$ is the mean of the distribution.
\end{itemize}

Both the Schechter and Gaussian input $P(\lambda)$ are normalized to unity. Such Eddington ratio distributions have a lower cutoff at $\lambda_{\min}$ = $10^{-4}$ below which the sources are no longer regarded as AGN\@. We choose $\lambda_{\min}$ to be low enough to include even the faintest AGN recorded in the $z=0.1$ AGN XLF, down to $\log L_X\sim 41$ erg/s for BHs with mass $\log M_\mathrm{BH} \gtrsim 6$. We note that the exact choice of $\lambda_{\min}$ is not too relevant in our modelling. By lowering/increasing $\lambda_{\min}$ would simply correspond to a higher/lower duty cycle, i.e.\ a higher/lower probability for BHs to active.
In Sec. \ref{sec6} we explore the effect on our results of assuming an input BH mass dependent Eddington ratio distribution, i.e. $P(\lambda,M_{bh})$.

We then assign a bolometric luminosity to each source according to the Eddington ratio $\lambda$ and the BH mass. The bolometric luminosity is then converted into intrinsic rest-frame 2-10 keV X-ray luminosity via the relation $L_X = L_{bol}/K_X$ with the bolometric correction $K_X$ expressed as
\begin{equation}
K_X(L_{bol}) = a \left[ 1 + {\left( \frac{\log(L_{bol}/L_{\odot})}{b} \right )}^c  \right]
\end{equation}
with a = 10.96, b = 11.93, c = 17.79
\citep[][]{duras20}.

\subsection{Input BH duty cycle}

In general terms, the total accretion probability
of a BH to be active at a given Eddington ratio
is a convolution of the duty cycle $U(M_\mathrm{BH})$,
i.e.\ the probability of a galaxy/BH to active as an
AGN above a certain luminosity threshold, and the
(normalized) Eddington ratio distribution $P(\lambda$)
of being accreting at a given rate
\citep[e.g.][]{steed03,marconi04,aversa15}.
We here follow the rather common and broad approximation
followed in the continuity equation formalism of
expressing the total accretion probability into a
simple product of the duty $U(M_\mathrm{BH})$
and the Eddington ratio distribution $P(\lambda)$
\citep[e.g.][]{small92,marconi04,shankar09}.
This choice is extremely flexible and allows to
disentangle the roles of a mass- and/or time-dependent
duty cycle from an evolving characteristic Eddington ratio $\lambda$
\citep[e.g.][]{shankar13}.

Both the Eddington ratio distribution $P(\lambda$)
and duty cycle $U(M_\mathrm{BH})$ have been separately
studied by different groups. For the duty cycle
in particular, despite the numerous dedicated works,
no clear trend has yet emerged and controversial
results are present in the literature. For this reason, we decided to test
three different duty cycles for BHs with mass
$\log(M_{\mathrm{BH}}/M_{\odot}) \ga 6$ as shown is Figure~\ref{fig3}:
\begin{itemize}
	\item A duty cycle $U(M_\mathrm{BH})$ decreasing with BH mass, as derived in
    \citet[][]{schulze10}
    at $z = 0.1$ for Compton thin un/obscured AGN\@, labelled as U-SW10 (decr) hereafter;
	\item A duty cycle increasing with BH mass in a way to reproduce the increasing trend with host galaxy stellar mass as estimated at low redshift (z$<0.1$) by
  \citet[][]{man19} for Narrow Line AGN in host galaxies with $\log(M_{\mathrm{star}}/[M_{\odot}])>9$, U-M19 (incr) hereafter;
\item A constant duty cycle $U(M_\mathrm{BH}\ \mathrm{or}\ M_{\mathrm{star}}) = 0.2$, as suggested by \citet[][]{goulding10}, U-G10 (const) hereafter.
\end{itemize}

In all cases we define as duty cycle the probability of BHs to be active above the minimal Eddington ratio threshold $\lambda_{\min}$ in our input $P(\lambda)$ distribution. It is worth noticing that we are assuming that the duty cycle from
\citet[][]{man19}
can be applied to both obscured and unobscured AGN\@. Given that it has been derived by using a sample of Narrow Line AGN, we can
consider it as a lower limit. However, a similar duty cycle increasing with
stellar mass has also been derived in
\citet[][]{georgakakis17}
from a sample of Compton thin un/obscured AGN\@.

\begin{table*}
\begin{center}
\begin{tabular}{lrrcrr}
  \toprule
  $M_{\mathrm{star}}-M_{\mathrm{BH}}$     & $U$ & $P(\lambda)$ & log$\lambda^{\star}$ (or log$\mu$) & $\alpha$ ( or $\sigma$) & log$\left \langle \lambda_{AR} \right \rangle$ \\
  \midrule
Sha16 (Debiased) & \citet[][]{schulze10}  & Schechter  & -0.45  & 0.15 & 31.81 \\
Sha16 (Debiased) & \citet[][]{man19}      & Schechter  & -1.8  & -0.15 &31.56 \\
Sha16 (Debiased) & \citet[][]{goulding10} & Schechter  & -1.9  & 0 & 31.56 \\
\midrule
SG16 (Raw) & \citet[][]{schulze10}  & Schechter  & -1.3  & -0.35 & 32.93 \\
SG16 (Raw) & \citet[][]{man19}      & Schechter  & -2.5  & 0.4  & 31.8 \\
SG16 (Raw) & \citet[][]{goulding10} & Schechter  & -2.4  & 0.8 & 31.42 \\
\midrule
Sha16 (Debiased) & \citet[][]{schulze10}  & Gaussian  & -2.8  & 1  & 31.69\\
Sha16 (Debiased) & \citet[][]{man19}      & Gaussian  & -4  & 1  & 31.51 \\
Sha16 (Debiased) & \citet[][]{goulding10} & Gaussian  & -4  & 1.1 & 31.24 \\
\midrule
SG16 (Raw) & \citet[][]{schulze10}  & Gaussian  & -3  & 1.1  & 32.62\\
SG16 (Raw) & \citet[][]{man19}      & Gaussian  & -4.5  & 1.3  & 31.71\\
SG16 (Raw) & \citet[][]{goulding10} & Gaussian  & -4.5  & 1.4  & 31.41\\
  \bottomrule
\end{tabular}
\caption{Model input parameters.}
\label{tab:edd}
\end{center}
\end{table*}

\subsection{The Q parameter}

The AGN duty cycle $U(M_\mathrm{bh})$ is the average fraction of both central and satellite galaxies to be active at a given stellar or BH mass, above a given threshold. However, the relative probability for a central and satellite BH to be active could still be different. To allow for this possibility, following
\citet[][]{shankar20}
we define the total duty cycle as the sum of central and satellite at a given BH mass to be active, i.e. $U(M_\mathrm{bh}) = U_c(M_\mathrm{bh}) + U_s(M_\mathrm{bh})$, with $U_c$ and $U_s$ the duty cycles of, respectively, satellite and central galaxies above a given luminosity or Eddington ratio threshold. We can then define the parameter $Q = U_s/U_c$ as the relative probability of satellite and central AGN of being active. Constraining the $Q$ parameter would be of course of key importance to shed light on the different AGN triggering mechanisms. For example, a high value of $Q$ would point towards satellites being preferentially active rather than centrals of similar mass, a condition that would be difficult to reconcile with a strict merger-only scenario but possibly still consistent with disc instability processes
\citep[e.g.][]{gatti16}.

Previous studies in the literature always assumed $Q$ = 1
\citep[noticeable exceptions are][]{allevato19,shankar20},
implying that all central and satellite
galaxies share equal probabilities of being active
\citep[e.g.][]{comparat19,aird20}.
The $Q$ parameter can in principle be directly measured from the fraction of satellite galaxies in groups and clusters of galaxies $f_{sat}^\mathrm{AGN}$
\citep[see][and references therein]{gatti16}.
In fact the $Q$ parameter can be expressed in terms
of $f_{sat}^\mathrm{AGN}$ as
$Q   = f_{sat}^\mathrm{AGN} (1-f_{sat}^{BH}) / [{1-f_{sat}^\mathrm{AGN}}]f_{sat}^{BH} $,
where $f_{sat}^{BH}=N_s/(N_s+N_c)$ is the total
fraction of (active and non active) BHs in satellites with BH mass within $M_{\mathrm{bh}}$ and $M_{\mathrm{bh}} + dM_{\mathrm{bh}}$
\citep[for full details, see][]{shankar20}.

\section{Outputs}

We then consider different outputs of our mock catalog of galaxies and BHs at a given z:
\begin{itemize}
	\item The AGN X-ray luminosity function (XLF):
	\begin{equation}\label{eq6}
	\begin{split}
    \Phi_\mathrm{AGN} (L_X) = & \int_{\log \lambda_{\min}} P(\lambda\propto L_X/M_\mathrm{bh}) \\
    & \times \: \mathrm{U}(M_\mathrm{bh}) \: \Psi (M_{\mathrm{bh}}) \: \mathrm{dlog} \lambda
    \end{split}
    \end{equation}
 \end{itemize}
     where $\Psi (M_{\mathrm{bh}})$ = $\Psi_\mathrm{AGN} (M_{\mathrm{bh}})/U(M_{\mathrm{bh}})$ is the total (active and non active) BH mass function, $U(M_{\mathrm{bh}})$ is the AGN duty cycle,
     $P(\lambda)$ is the normalized Eddington ratio distribution with $\log \lambda_{\min} = -4$.
\begin{itemize}
	\item The specific accretion rate distribution:
	\begin{equation}\label{eq7}
	\begin{split}
	P_\mathrm{AGN} (\lambda\propto L_X/M_{\mathrm{star}}) = & \int_{\log\lambda_{\min}} P(\lambda\propto L_X/M_{\mathrm{star}}) \\ \: & \times \mathrm{U}(M_{\mathrm{star}}) \: \mathrm{dlog} \lambda
	\end{split}
	\end{equation}
 \end{itemize}
where $\lambda \propto L_X/M_{\mathrm{star}}$ defines the rate of accretion onto the central BH scaled relative to the stellar mass of the host galaxy. $P_\mathrm{AGN}$ describes the probability of a galaxy to host an AGN of a given
$L_X/M_{\mathrm{star}}$ at a given redshift.
We can also define the characteristic
$\left \langle \lambda \right \rangle$ of the specific accretion-rate distribution as:
\begin{equation}
\left \langle \lambda \right \rangle = \frac{\int \lambda \: P_\mathrm{AGN} (\lambda) \: \mathrm{dlog} \lambda}{\int P_\mathrm{AGN} (\lambda) \: \mathrm{dlog} \lambda}
\end{equation}
At variance with many other previous approaches, our flexible methodology based on an input duty cycle and Eddington ratio distribution allow us to use the $P_\mathrm{AGN}$ distribution as an output rather than an input of our AGN mock catalog, thus providing an additional valuable constraint independent of AGN clustering. We will show that the $P_\mathrm{AGN}$ distribution is particularly useful in constraining the viable duty cycles and also the underlying BH-galaxy scaling relations.


\begin{itemize}
    \item The large-scale bias of mock AGN with BH mass (and similarly stellar mass) in the range $\log M_{\mathrm{bh}}$ and $\log M_{\mathrm{bh}} + d\log M_{\mathrm{bh}}$ following the formalism of
    \citet[][]{shankar20}:
 \end{itemize}


\begin{equation}
\begin{split}
  b =
  & \left[ \sum_{i=1}^{N_{c}} U_{c,i}(M_{\mathrm{bh}}) b_{c,i}(M_{\mathrm{bh}}) \right. \\
  & + \left. \sum_{i=1}^{N_{s}} U_{s,i}(M_{\mathrm{bh}}) b_{sat,i}(M_{\mathrm{bh}}) \right] \\
  & \left/ \left[ \sum_{i=1}^{N_{cen}} U_{c,i}(M_{\mathrm{bh}}) + \sum_{i=1}^{N_{s}} U_{s,i}(M_{\mathrm{bh}}) \right] \right.
\end{split}
\label{eq9}
\end{equation}
where $U_{c}(M_{\mathrm{bh}}) = U(M_{\mathrm{bh}})N(M_{\mathrm{bh}})/(N_c(M_{\mathrm{bh}}) + Q N_s(M_{\mathrm{bh}}))$ is the duty cycle of central AGN, $U_{s}(M_{\mathrm{bh}}) = Q U_{c}  (M_{\mathrm{bh}})$ is the duty cycle of satellite AGN
and $N(M_{\mathrm{bh}}) = N_c(M_{\mathrm{bh}}) + N_s(M_{\mathrm{bh}})$ is the number of central and satellite galaxies,
in the BH mass bin $M_{\mathrm{bh}}$ and
$M_{\mathrm{bh}} + dM_{\mathrm{bh}}$. 

\begin{figure*}
\centering
\plottwo{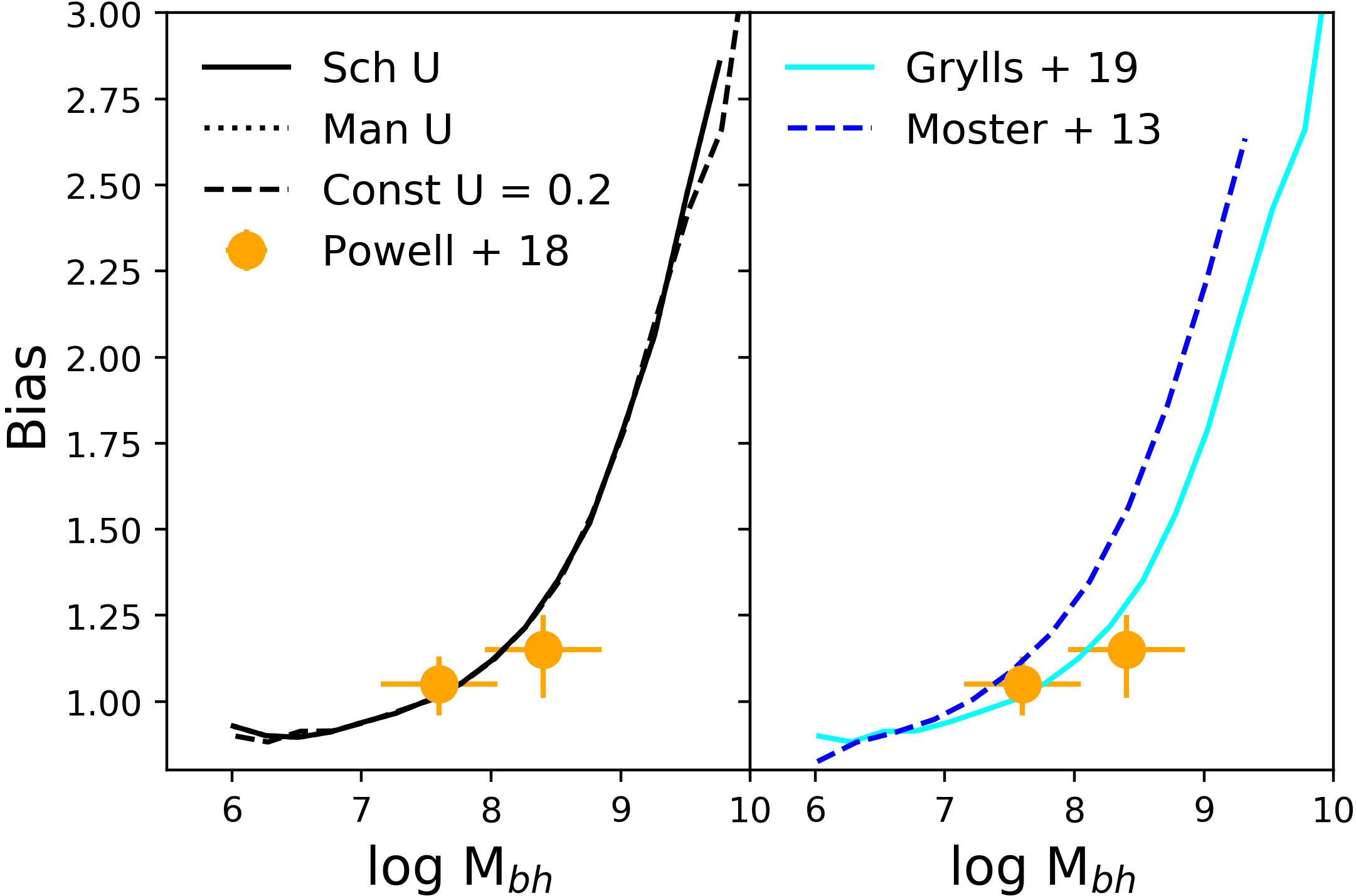}%
        {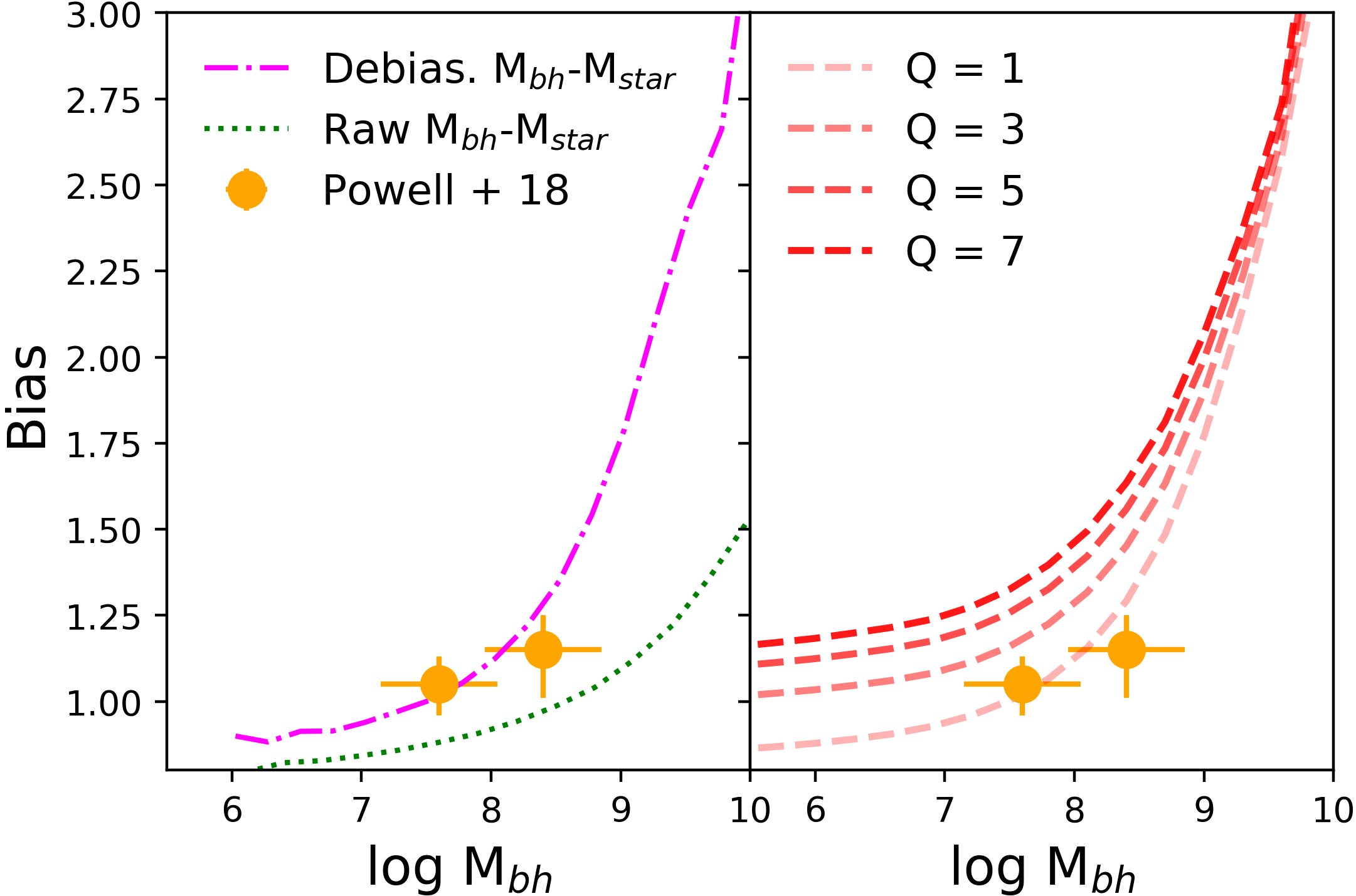} \\
\plottwo{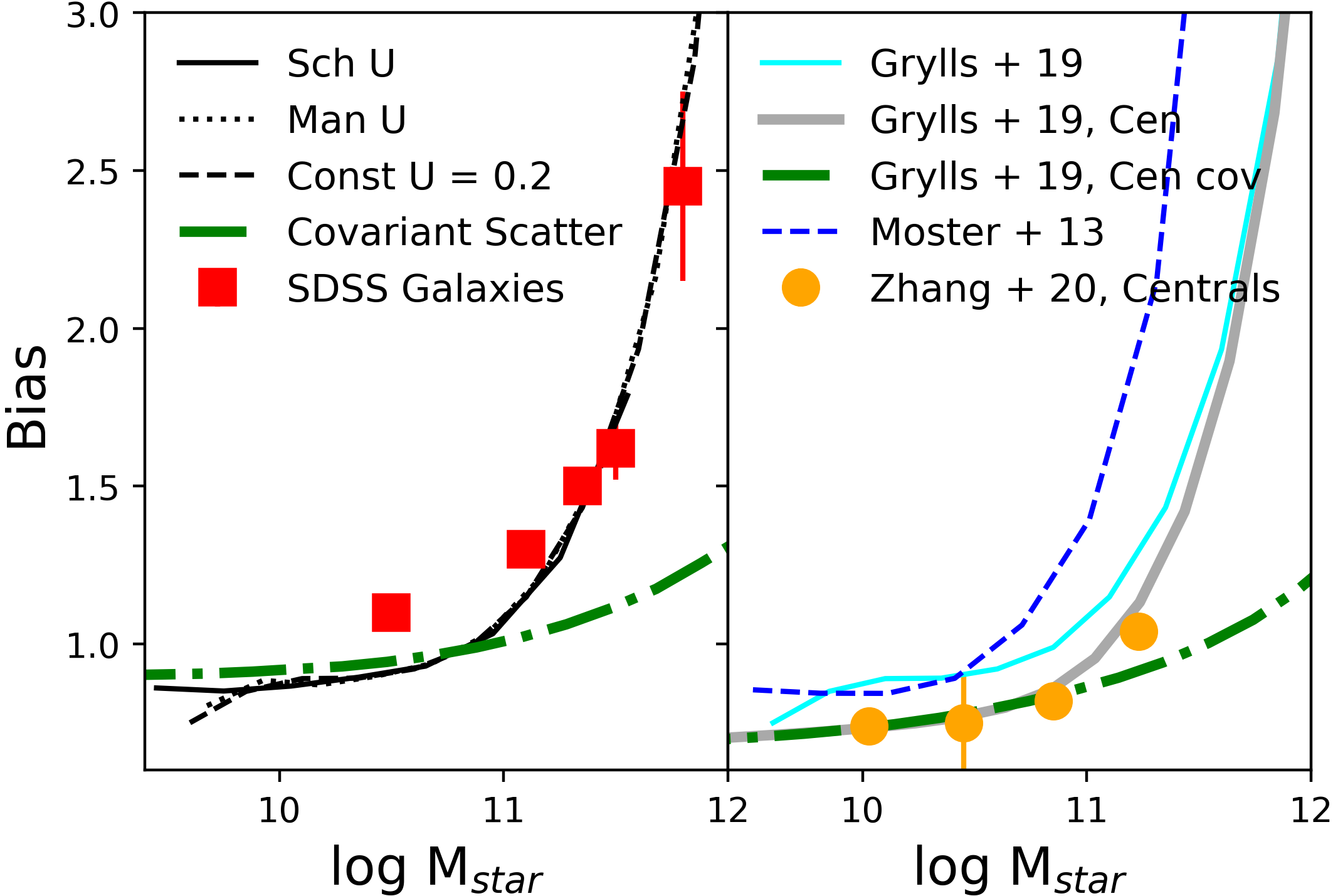}%
        {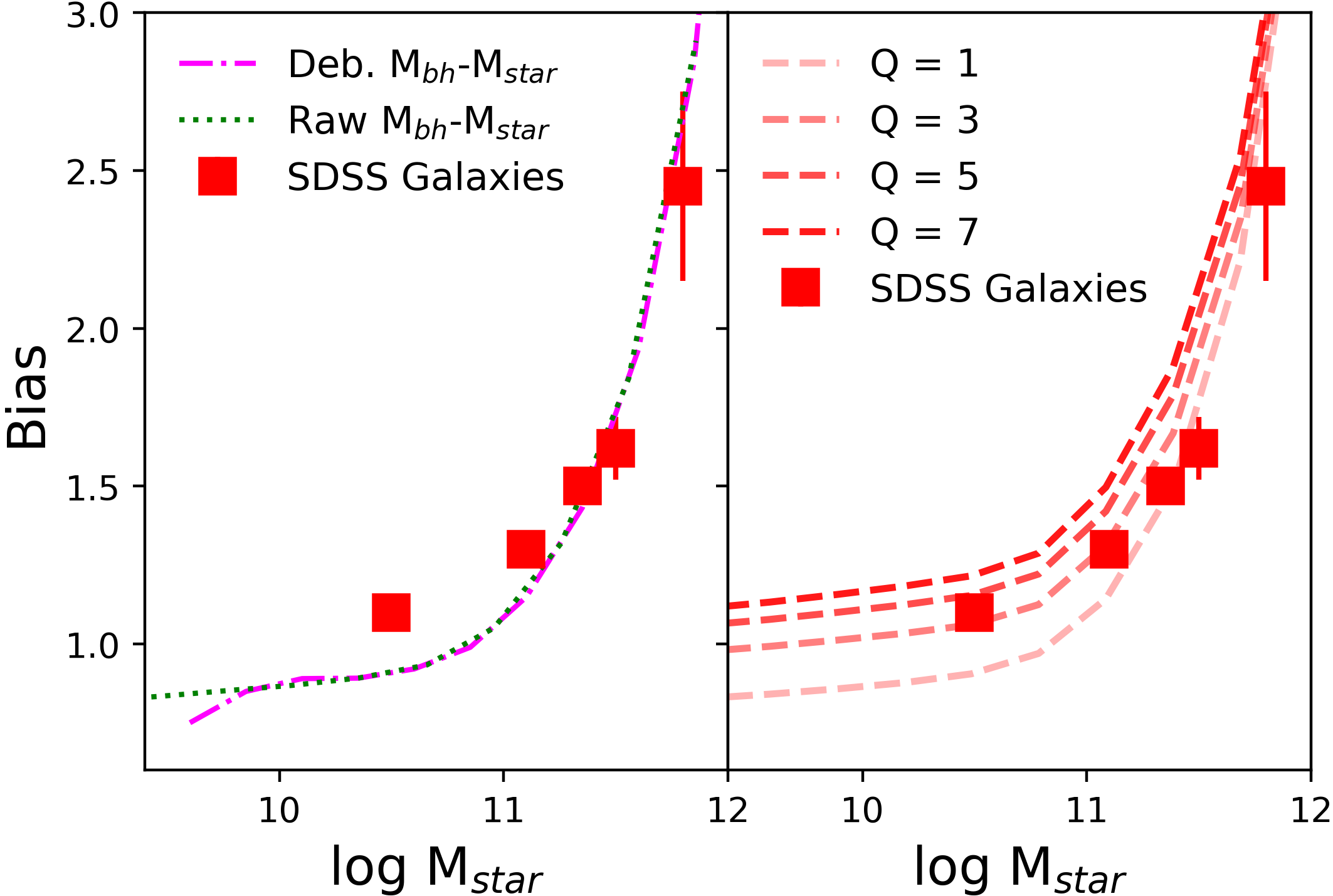}%
\caption{\footnotesize Large-scale bias of mock AGNs as a function of BH mass (upper panel) and host galaxy stellar mass (lower panel), when using different duty cycles, input stellar mass--halo mass and BH mass--stellar mass relations and $Q$ (see text for more details). The bias estimates as a function of $M_{\mathrm{BH}}$ from
\citet[][]{powell18} for X-ray selected AGN at z $\sim$ 0.04, as a function of $M_{\mathrm{star}}$ from SDSS AGN 
\citep[][orange circles]{zhang20} and SDSS galaxies \citep[][red squares, see the text for more details]{dominguez18} in the local Universe are shown for comparison. The green dash-dot line shows the predictions when assuming a covariant scatter in the $M_{star}-M_h$ and $M_{bh}-M_{star}$ relations as discussed in Sec~\ref{sec6}.} 
\label{fig8}
\end{figure*}

\begin{figure*}
	\plottwo{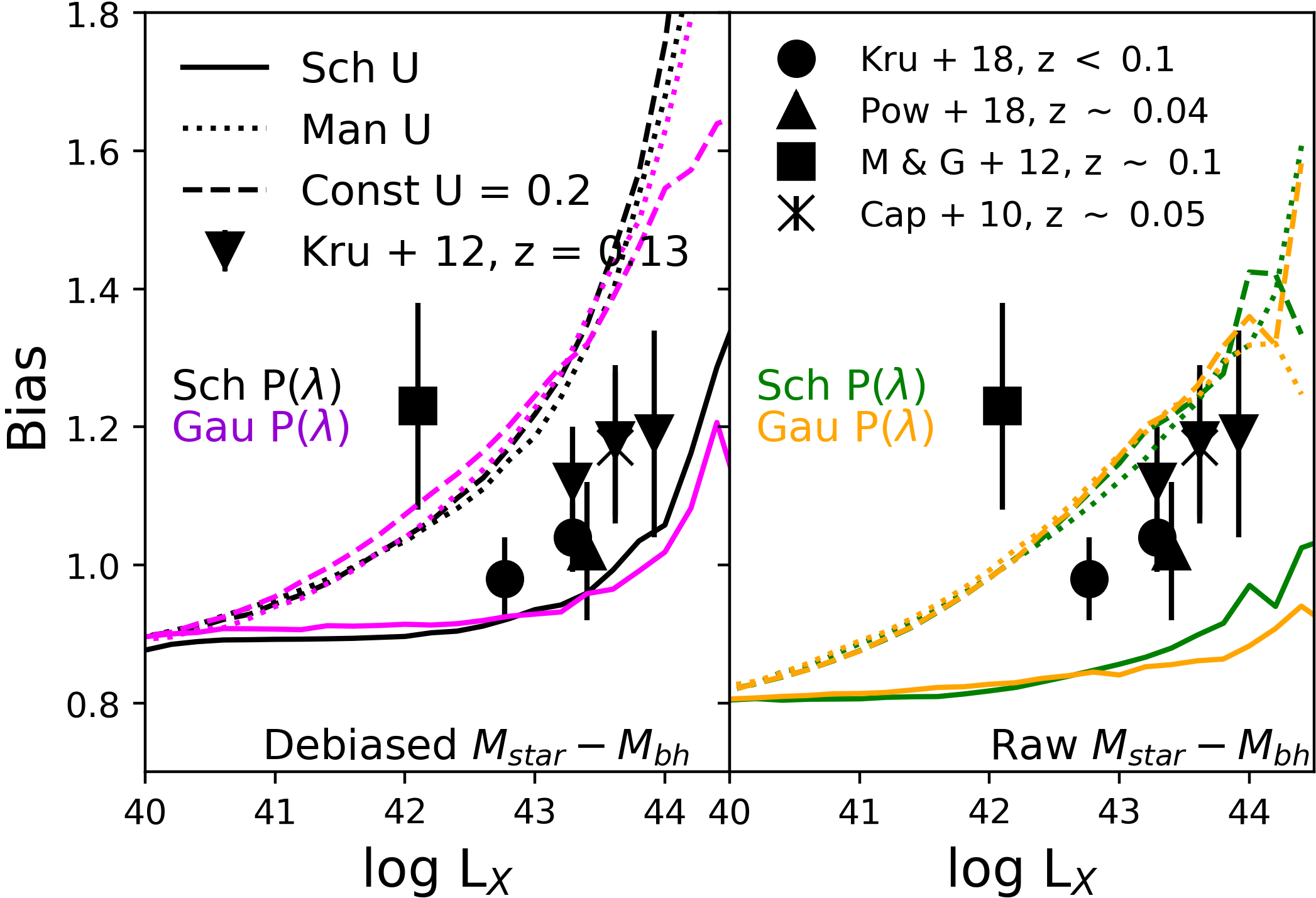}{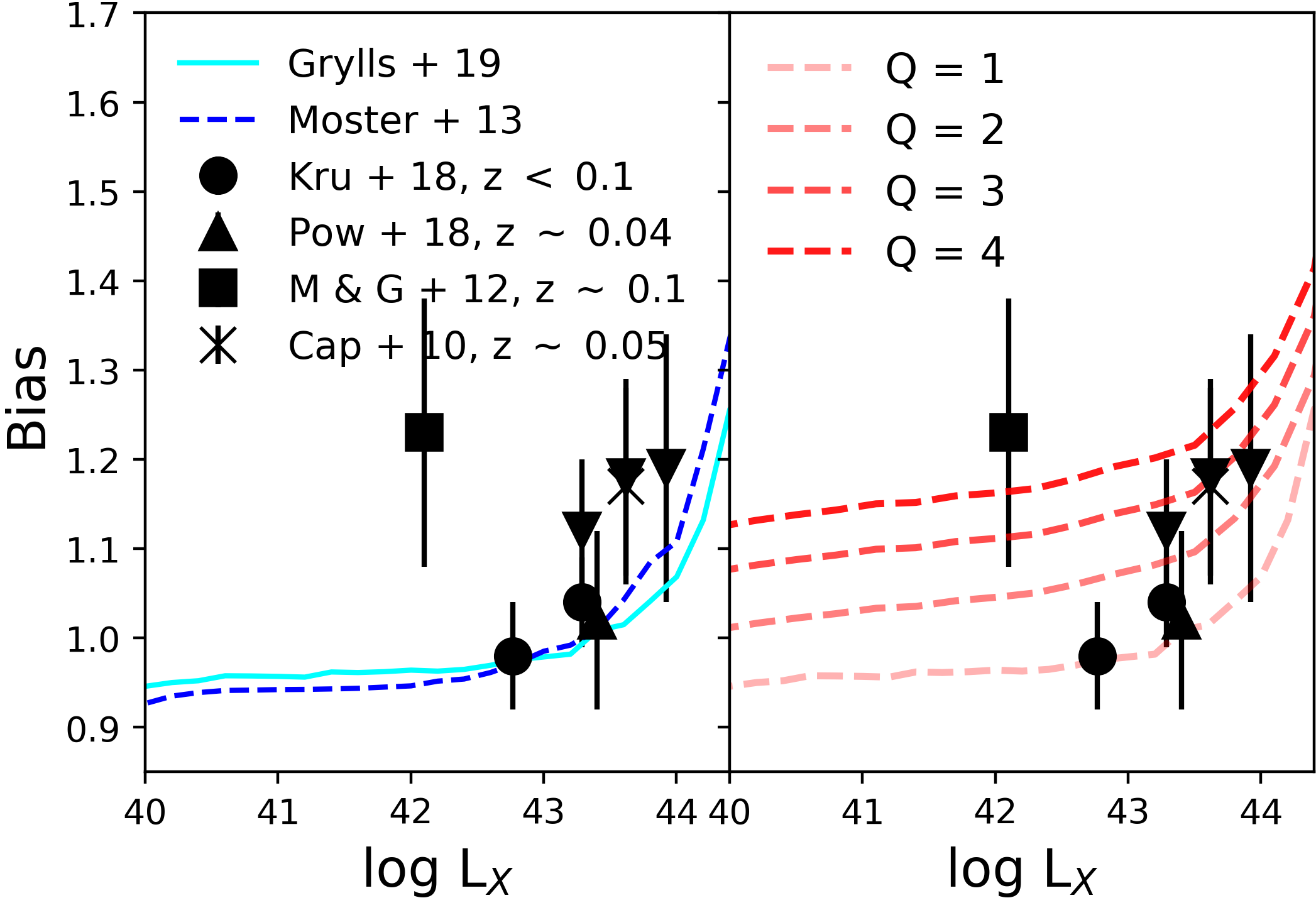}
 	\caption{\footnotesize Large-scale bias versus X-ray (2-10 keV) luminosity (in units of erg/s) for mock AGN when using a decreasing \citep[][]{schulze10},
increasing \citep[][]{man19}, and constant \citep[$U = 0.2$,][]{goulding10}
duty cycle as a function of BH mass, compared with bias estimates from previous studies at similar redshifts. For
samples for which the AGN X-ray luminosity is estimated in an energy interval other than the 2–10 keV band, L$_X$(2--10 keV) is derived assuming a power-law X-ray spectrum with photon index of $\Gamma$ = 1.9}
\label{fig10}
\end{figure*}

\section{Results}

\subsection{AGN XLF and \texorpdfstring{$P_\mathrm{AGN}$}{PAGN}}

We now use the model described in the previous section to create mock
catalogs of AGN\@. We vary the input parameters and study how these changes
affect the outputs, such as the AGN XLF and the specific accretion-rate distribution $P_\mathrm{AGN}$.
We consider various different cases, and each combination of model input parameters are shown in Table~\ref{tab:edd}.

We estimate the XLF $\Phi_\mathrm{AGN}$ (defined in
equation~\ref{eq6}) for the different duty cycles,
setting the free parameters
of the input $P(\lambda$) distribution in order to reproduce the
observationally inferred XLF of X-ray selected AGN at z = 0.1
\citep[][]{miyaji15}.
We infer the parameters of $P(\lambda$) based on
an overall match to an observational constraint, i.e.\
the AGN XLF, but we do not attempt formal
$\chi^2$- minimization because the errors
themselves are not well defined enough to do so.
We note where models fit observations within
a plausible range of systematic uncertainties, and where they do not.
Our objective is to show how the different
`observables' depend on the input model parameters
and delineate a guideline for the creation
of realistic AGN mock catalogs, so we are not inherently
interested in the free parameters of $P(\lambda$)
that might better fit the (real) observations.

As a first case, we consider a $P(\lambda$) described by a Schechter function. As shown in Figures~\ref{fig4} and~\ref{fig5}, we can reproduce the observed AGN XLF as derived in \citet[][]{miyaji15}
for AGN at z = 0.1
\citep[or similarly in][]{ueda14},
independently of the choice of the input AGN
duty cycle and BH mass--stellar mass relation.
In particular, if we assume BHSMR-Sha16 (debiased),
the input $P(\lambda$) is characterised by a power-law
index $\alpha \sim 0$, i.e.\ an almost constant probability
as a function of Eddington ratio at $\lambda < \lambda^{\star}$.
Assuming the AGN duty cycle U-SW10 (decr),
the input $P(\lambda$) is almost consistent with observations
\citep[i.e.][]{Kauffmann09,hickox09}
with a knee log$\lambda^{\star}$ = - 0.45.
It is worth noting that these observations are calibrated
on the $M_{BH}-\sigma$ relation of Tremaine et al. (2002) 
that would shift the Eddington ratio distribution to higher $\lambda$ 
by a factor of $\sim$2,
still in agreement with our
AGN mock predictions. A smaller knee
(log$\lambda^{\star} \sim$ -2) is obtained for mock AGN
assuming U-M19 (incr) and U-G10 (const).
Similarly, we can reproduce the observed AGN XLF for
different AGN duty cycles when assuming SG16 (raw)
(see figure~\ref{fig5}). However, the corresponding
input $P(\lambda$) distributions of mock AGN with U-M19 (incr)
and U-G10 (const) are in tension with the data at similar redshifts
\citep[i.e.][]{Kauffmann09,hickox09}.

We also derived the corresponding specific accretion-rate distributions $P_\mathrm{AGN}$ defined in equation~\ref{eq7} as the convolution of an input normalized $P(\lambda$) and the AGN duty cycle $U$. We also applied to mock AGN a luminosity cut at $\log(L_X/[\mathrm{erg}\,\mathrm{s}^{-1}]) > 41$ in order to compare with recent observations based on X-ray selected AGN\@. As shown in figure~\ref{fig6} and~\ref{fig7}, $P_\mathrm{AGN}$ is affected by
the input $P(\lambda$) and the duty cycles (for a given BH mass--stellar mass relation) and by the $M_{\mathrm{star}}-M_{\mathrm{BH}}$ relation (for a given duty cycle).
It is immediately noteworthy that the specific accretion-rate distribution mimics the shape of the input $P(\lambda$) while the AGN duty cycle
affects the characteristic $\left \langle \lambda \right \rangle$. In detail,
assuming BHSMR-Sha16 (debiased)
and setting U-SW10 (dec),
the $P_\mathrm{AGN}$ distribution is more
consistent with observations at similar redshifts
\citep[e.g.][]{aird17,georgakakis17}
and has a larger characteristic
log$\left \langle \lambda \right \rangle$ (= 31.8 and 31.7 for a Schechter and Gaussian $P(\lambda$), respectively) than using different $U$ (see Table~\ref{tab:edd}). However, the observationally
derived specific accretion-rate distributions are characterized by tails at high $L_X/M_{\mathrm{star}}$ that are not present in our mock AGN predictions (see Section~\ref{sec6} for more discussion).

When we use BHSMR-SG16 (raw), the $P_\mathrm{AGN}$
distribution of mock AGN is almost one order of magnitude higher at all
$L_X/M_{\mathrm{star}}$ than data, when using U-SW10 (decr). The $P_\mathrm{AGN}$ distribution of mock AGN is more in line with observations when using
U-M19 (incr) and U-G10 (const), at least at log$L_X/M_{\mathrm{star}} \le 33$. However, the corresponding input Eddington ratio distributions $P(\lambda$) are highly inconsistent with observations (see fig.~\ref{fig5}).

It is worth noting that all these results are
derived for a $M_{\mathrm{star}}-M_{h}$ relation given by
\citet[][]{grylls19} and are independent of the particular choice of the $Q$ parameter.

\subsection{AGN Large-scale bias}

Each mock AGN resides in satellite or central halos
with a given parent halo mass that corresponds to a
specific value of the large-scale bias via the numerically-derived correlation between halo mass and bias which we take from \citet[][]{vandenbosch02} and \citet[][]{tinker05}, in line with what assumed in the observational samples.
We then derive the bias of mock AGN as a function of the
host galaxy stellar mass and BH mass by using Eq.~\ref{eq9}
with different choices of the underlying duty cycles $U$,
input stellar mass--halo mass, BH mass--stellar mass relations,
and values of the $Q$ parameter. In all the model renditions considered below, the $P(\lambda$) parameters are fixed in order to reproduce the AGN XLF\@.

Figure~\ref{fig8} shows the AGN large-scale bias as a function of BH mass and host galaxy stellar mass, when using different (a) duty cycles (for fixed $M_{\mathrm{star}}-M_h$ and $M_{\mathrm{bh}}-M_{\mathrm{star}}$ relations and $Q$); (b) input stellar mass--halo mass relations (for fixed $M_{\mathrm{bh}}-M_{\mathrm{star}}$ relation, duty cycle and $Q$); (c) input BH mass--stellar mass relation (for a fixed duty cycle,$M_{\mathrm{star}}-M_{h}$ relation and $Q$); (d) $Q$ values (for fixed duty cycle and $M_{\mathrm{star}}-M_{h}$ and $M_{\mathrm{bh}}-M_{\mathrm{star}}$ relations).

As expected, the large-scale bias as a function of BH mass mainly depends on the input BH mass--stellar mass relation and $Q$ parameter, with a mild dependence on the $M_{\mathrm{star}}-M_{h}$ relation (top panels of Figure~\ref{fig8}). Conversely, the bias as a function of the AGN host galaxy stellar mass is only affected by $Q$, with a weak dependence on the input stellar mass--halo mass relation (bottom panels of Figure~\ref{fig8}). In particular, mock AGNs with a given BH mass reside in more massive parent halos when assuming BHSMR-Sha16 (debiased) and/or $Q \ga 2$. In the former case, the effect is stronger at large BH masses, while in the latter is mainly affecting small BH masses. These results are independent of the shape of the input $P(\lambda)$ distribution, either Gaussian or Schechter.

The comparison of our model predictions with large-scale
bias estimates of X-ray selected AGN as a function of
$M_{\mathrm{BH}}$ in the local Universe
\citep[][]{powell18}
show a degeneracy among the input model parameters.
In fact, the observations can be reproduced by either
BHSMR-Sha16 (debiased) with $Q$ = 1
\citep[using][]{grylls19};
or by a BHSMR-SG16 (raw) with $Q > 2$ and/or a
stellar mass--halo mass relation given by
\citet[][]{moster13}.
It is worth noting that the BH masses in
\citet[][]{powell18}
are derived by parameters calibrated
on relations close to BHSMR-SG16 (raw). A better
comparison with a model that assumes BHSMR-Sha16 (debiased) would imply a correction that moves the data to lower BH masses, strengthening the agreement among the observations and our model predictions.

Unfortunately, only few measurements are available at z $\le$ 0.1 of the AGN large-scale bias in bins of host galaxy stellar mass. In particular, recent estimates of the hosting central halo mass of SDSS AGN
\citep[][]{zhang20}
  suggest a $M_{\mathrm{star}}-M_{h}$ relation in agreement with
\citet[][]{grylls19}.
To provide additional clustering constraints, we derived the 2-point projected correlation function $w_p(r_p)$ in the range r$_p = 0.1 - 30 \; h^{-1} Mpc$, as a function of stellar mass for the SDSS galaxies at z $<$ 0.1 \citep{dominguez18}. This sample has the same photometry and mass-to-light ratios as those adopted in the \citet{bernardi17} stellar mass function which we adopt as a reference for our stellar mass-halo mass relation (Figure~\ref{fig1}). We then converted $w_p(r_p)$ to bias estimates by making use of the projected 2-point correlation function of the matter \citep{EisenHu} with the same cosmology as in our reference dark matter simulation. The results are shown as red squares in the bottom left panel of Figure~\ref{fig8}. The errors on the SDSS galaxies clustering measurements correspond to the square root of the covariance matrix diagonal elements calculated via the bootstrap resampling method.

Our predicted bias as a function of stellar mass nicely lines up with the SDSS galaxy bias measurements, especially for galaxies with mass $\log M_{\star}/M_{\odot} \gtrsim 11$ and, as expected, in ways fully independent of the duty cycle and the input BH mass--stellar mass relation. Our results thus strongly suggest that AGN mocks where the AGN activity is independent of environment (i.e. $Q \sim 1$), will guarantee a match to the galaxy clustering if the host galaxies are already tuned against clustering measurements (we discuss possible caveats to this statement in Section~\ref{sec6}).

Figure~\ref{fig7} shows that, despite the bias as a function of BH mass and galaxy stellar mass being excellent observables to constrain the BH mass--stellar mass relation and the $Q$ parameter, they are insensitive to AGN duty cycle. We discuss below how the AGN bias as a function of AGN luminosity can help to break the degeneracies in this fundamental input parameter.

As shown in Figure~\ref{fig10}, the AGN bias as a function of $L_{x}$ depends in fact mostly on the AGN duty cycle and $Q$ parameter (outer left and right panels), moderately on the BH mass--stellar mass relation (left panels), and weakly on the stellar mass--halo mass relation (inner right panel). The trends reported in the left panels of Figure~\ref{fig10} can be readily understood from the fact that an increasing duty cycle with BH mass (such as the U-M19, dotted lines) necessarily implies, on average, lower Eddington ratios to reproduce the same luminosity function, as proportionally more massive BHs will be active in this model. In turn, lower Eddington ratios will map the same AGN luminosities to more massive BHs residing, on average, in more massive and more clustered galaxies and dark matter halos. At fixed duty cycle and Eddington ratio distribution, a lower normalization in the BH mass--stellar mass relation, such as in our BHSMR-Sha16 (debiased) case, would map the same luminosities to more massive/clustered galaxies.

At face value, the comparison with the large-scale AGN bias as a function of $L_X$ estimated for X-ray selected AGN at z $\le$ 0.1 \citep[e.g.][]{krumpe18,powell18}, favours models adopting the BHSMR-Sha16 (debiased) relation and decreasing duty cycles, as in our U-SW10 (decr) model (left panels), in ways largely independent of the shape of the input $P(\lambda)$ distribution. We note that the data could also be reproduced by assuming BHSMR-SG16 (raw) and U-SW10 (decr) combined with $Q >$ 3, as this model would boost the clustering signal at all AGN luminosities due to a significant increase in the relative fraction of satellites, hosted in more massive/clustered parent halos, to be active (outermost right panel). However, this same model would also predict a $P_\mathrm{AGN}$ distribution an order of magnitude higher than the observationally derived specific accretion-rate distributions (see Figure~\ref{fig7}), while models based on the BHSMR-Sha16 (debiased) relation and the U-SW10 (decr) duty cycle would be consistent with it (see Figure~\ref{fig6}).


\begin{figure*}
	\plottwo{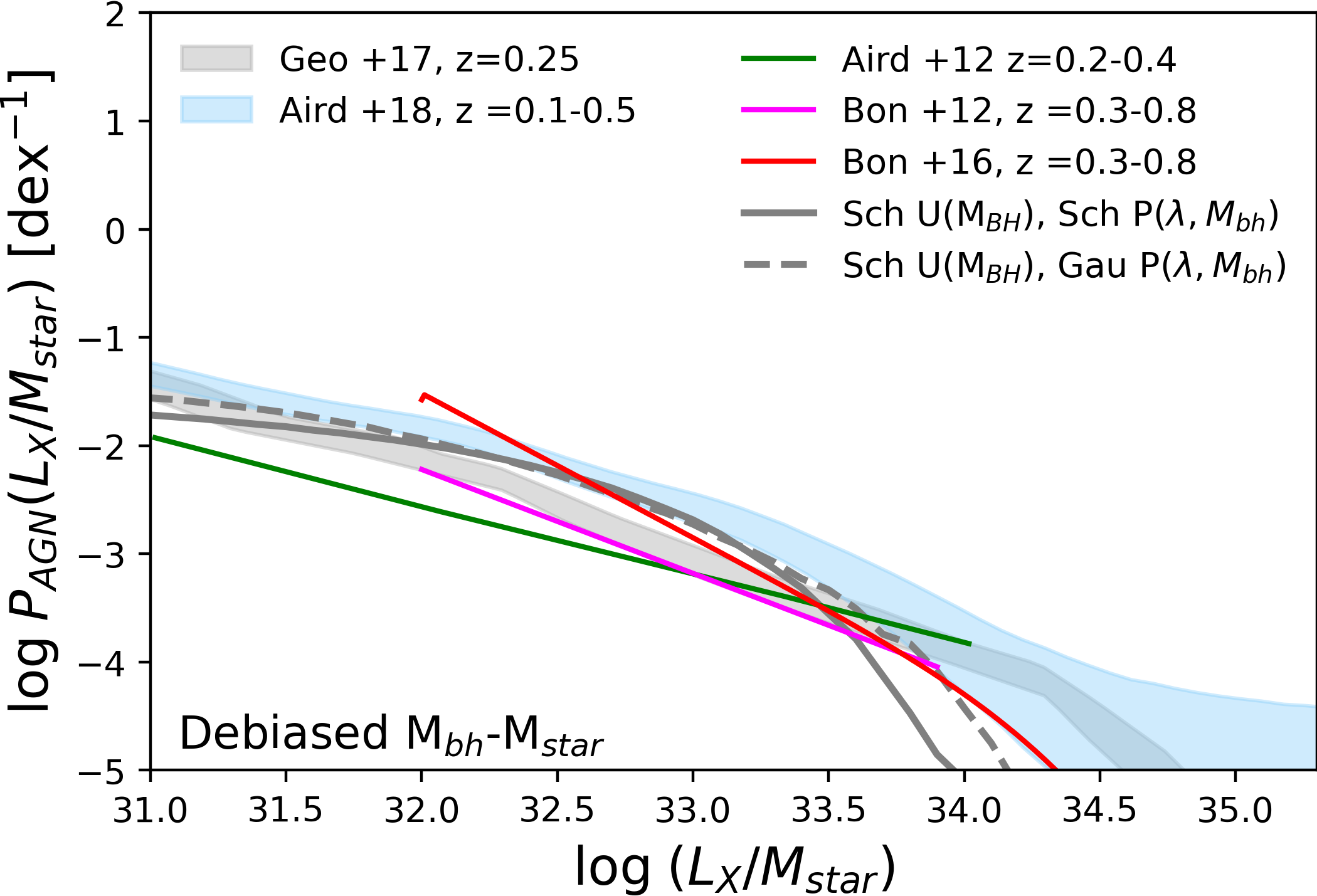}{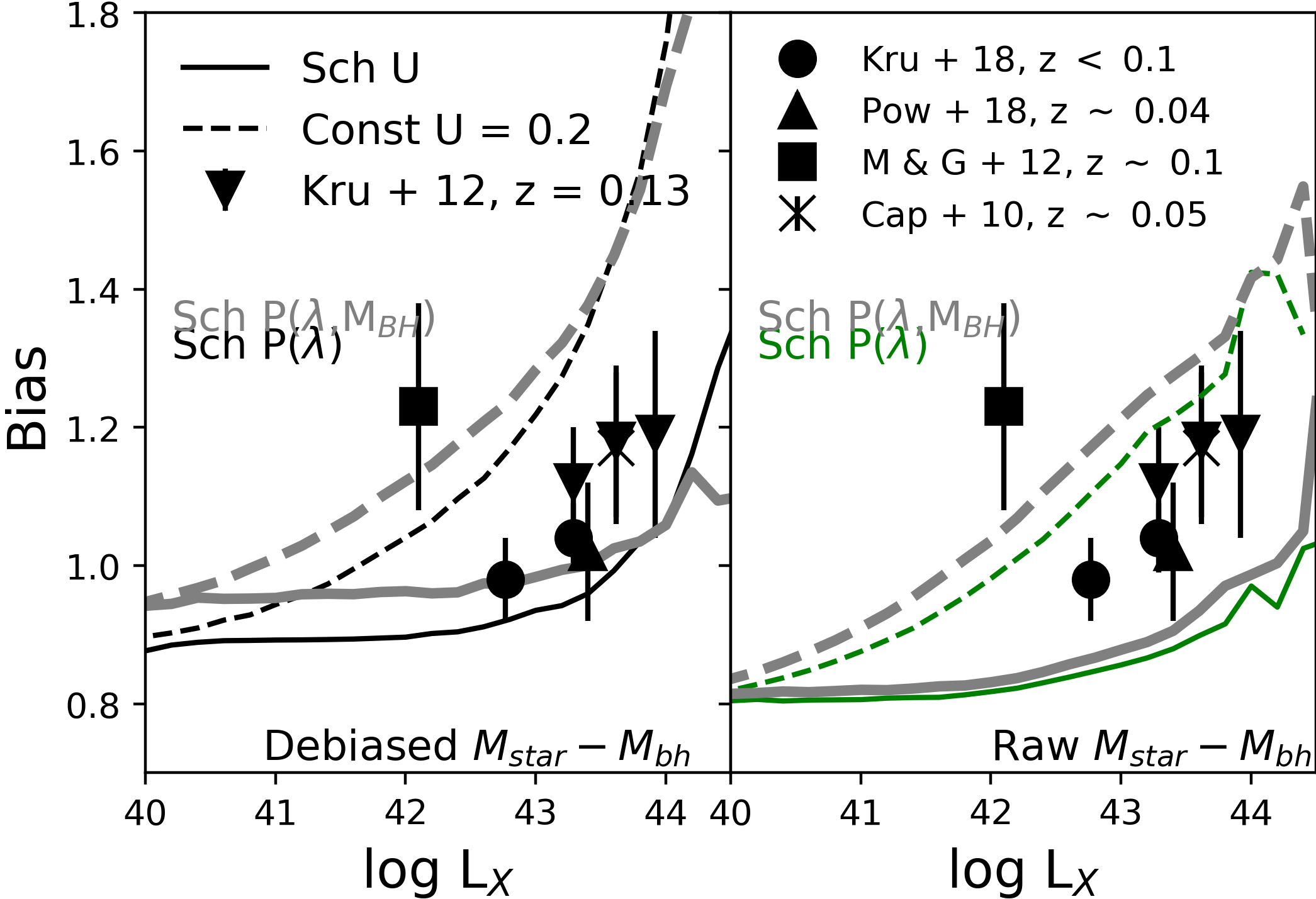}
 	\caption{\footnotesize Left panel: Specific accretion-rate distribution $P_{AGN} (\lambda, M_{star})$ given by the convolution of the input BH mass dependent Eddington ration distribution $P(\lambda,M_{bh})$ and AGN duty cycle. The prediction from mock AGNs with a Schecther (continuous line) and Gaussian (dashed line) $P(\lambda,M_{bh})$ and a $M_{star}-M_{BH}$ relation as defined in \citet{shankar16} are compared with data as in Fig.~\ref{fig5}. Right panel: Large-scale bias versus X-ray (2-10 keV) luminosity (in units of erg/s) for mock AGN when using an input Eddington ratio distribution that is independent (dependent) of the BH mass, for a decreasing \citep[][]{schulze10} and constant \citep[$U = 0.2$,][]{goulding10}
duty cycle as a function of BH mass, compared with bias estimates from previous studies at similar redshifts.}
\label{fig13}
\end{figure*}

\section{Additional dependencies in the input model parameters}\label{sec6}

All the reference models discussed so far to create AGN mock catalogs assume that the input parameters are uncorrelated. In this Section we explore the impact of relaxing this assumption in some of the key parameters in our modelling. As a first case, we assume that the BH, stellar and halo mass share some degree of correlation. 
More specifically, we assume that there exists a covariant scatter in the input stellar mass--halo mass relation and in the 
BH mass--stellar mass relation. 
In practice, we assign to each halo mass a value of M$_{star}$ and 
M$_{bh}$ from a multivariate Gaussian distribution following the methodology described in Viitanen et al. (submitted). A positive covariance would imply that it would be more likely for M$_{bh}$ to be scattered in the same direction as M$_{star}$.

We find that the covariance scatter does not affect the AGN large-scale bias 
as a function of BH mass and X-ray luminosity. On the contrary, as shown in Figure~\ref{fig8} (left panel), 
the AGN bias dependence on host galaxy stellar mass is smoothed out when assuming a covariant scatter, independently of the particular choice of the input stellar mass--halo mass and BH mass--stellar mass relation or AGN duty cycle. This behaviour is expected as the end effect of a covariant scatter is to generate a larger scatter in the scaling relations, thus naturally reducing the clustering strength especially at larger stellar masses. In particular, the covariant scatter model
predicts an AGN bias versus M$_{star}$ almost constant, and  
at $M_{star} \sim 10^{11.5} M_{\odot}$ two times smaller than what predicted by the case without covariance. A model with covariant scatter, i.e. with a (positive) correlation between BH mass and galaxy mass at fixed halo mass, would then imply a bias as a function of the stellar mass substantially different, at large stellar masses M$_{star} > 10^{11} M_{\odot}$, between AGN and the overall population of galaxies. In other words, a covariant scatter would inherently imply that AGN host galaxies are not a random selection of the galaxies of the same stellar mass. Present data do not allow us to clearly distinguish between models with and without a covariant scatter. 
In fact, as shown in Figure~\ref{fig8} (left panel), currently available 
AGN bias estimates as a function of stellar mass of SDSS AGN \citep{zhang20} only extend up to $M_{star} \sim 10^{11.3} M_{\odot}$, where the models have just started to diverge (solid gray versus long dashed green lines), although the data tend to be closer to the model without covariant scatter. AGN clustering measurement at higher host galaxy stellar mass bins will become available in the near future (e.g., 
Euclid) allowing us to rule out, or confirm, a covariant scatter at a high confidence level. We stress that, as anticipated above, the model without covariance is in good agreement, as expected, with the bias of SDSS galaxies at z$<$0.1 (red squares in Figure~\ref{fig8}, outermost left panel). We will more comprehensively discuss the consequences of a covariant scatter in Viitanen et al. (submitted).
\begin{figure*}
	\plotone{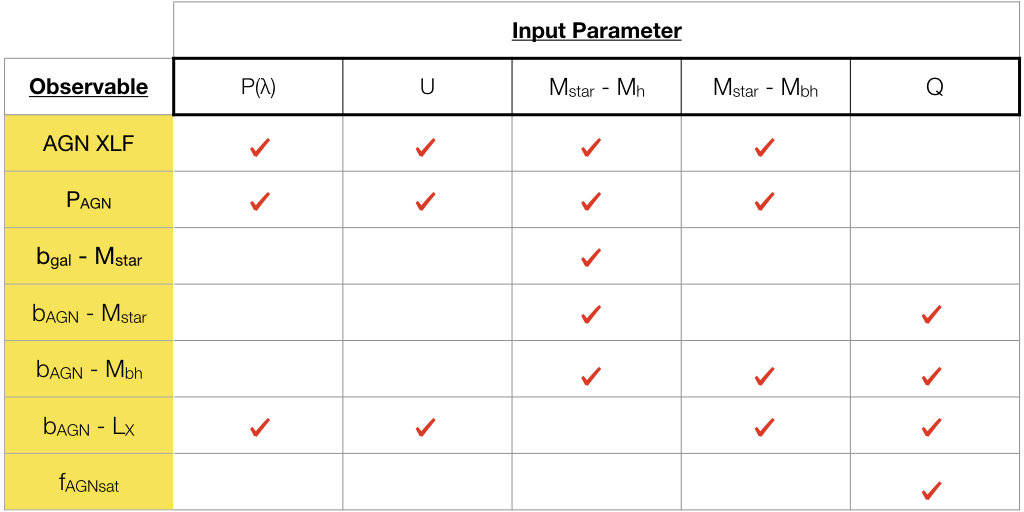}
 	\caption{\footnotesize Dependence of the observables on the input model parameters.}\label{fig11}
\end{figure*}

As a second relevant case we explore the effect of using an input 
Eddington ratio distribution P($\lambda$) that also depends on the BH/stellar mass, 
i.e. P($\lambda,M_{bh}$). A mass dependence in the input Eddington ratio distribution is expected from, e.g., continuity equation arguments \citep[e.g.,][]{shankar13,aversa15}, as well as from direct observational measurements \citep[e.g.,][]{Kauffmann09,georgakakis17,aird18}. 
Broadly following the continuity equation model by \citet[][see their Figure~6]{aversa15}, we divide our BH mock sample into two groups above and below a dividing mass of logM$_{bh}$ [M$_{\odot}$] = 7, and then assign to each group of BHs Eddington ratios extracted from a Schechter P$(\lambda)$ with the same power-law slope $\alpha$ and a higher characteristic $\left \langle \lambda \right \rangle$ for BHs in the lowest mass bin.

As discussed above, we expect that varying the input Eddington
ratio distribution will mainly affect two observables, namely the specific accretion-rate distribution P$_{AGN}$ and the AGN large-scale bias as a function of X-ray luminosity. 
Indeed, we verified that all our main predictions remain unaltered when adopting a mass-dependent P($\lambda,M_{bh}$, and found only a moderate variation in P$_{AGN}$ (left panel of Figure~\ref{fig13}), with a more pronounced tail at higher $L_X/M_{star}$, in somewhat better agreement with the data. Similarly, we find that an input P($\lambda,M_{bh}$) only slightly increases the AGN bias vs L$_X$ by $\sim 5\%$, compared to an input Eddington ratio distribution independent of BH mass (right panel of Figure~\ref{fig13}).

All in all, from the tests discussed above we can conclude that introducing reasonable correlations among the main parameters at play in our model does not significantly alter any of our main results.


\section{How to build realistic AGN mocks}

In the previous sections we showed that a large variety of models
characterised by distinct $M_{\mathrm{BH}}-M_{\mathrm{star}}$ relations and specific accretion-rate distribution $P_\mathrm{AGN}$ (obtained as convolution of the input $P(\lambda$) with the AGN
duty cycle $U$), can create AGN mocks matching the observed AGN XLF\@. In addition, the corresponding large-scale bias at a given stellar mass is independent of $P_\mathrm{AGN}$ and the stellar mass--BH mass relation, simply because the bias mostly depends on the parameter $Q$
and the input $M_{\mathrm{star}}-M_h$ relation.
Thus, having characterised a given $P_\mathrm{AGN}$ that, by design, observationally fits the AGN XLF, does not guarantee a unique and valid model to create AGN mocks even when we consider the clustering at fixed stellar mass, simply because the latter is not affected by the $P_\mathrm{AGN}$ distribution and the stellar mass--BH mass relation.

The results summarised above imply strong degeneracies among the input parameters used to create mock catalogs of AGN\@. Only considering all the observables,
in particular the AGN large-scale bias as a function of both BH mass and
X-ray luminosity, we can break such degeneracies in the input model parameters. Figure~\ref{fig11} provides a table summary of the different dependencies of the observables considered in this work on one or more of the model input parameters. Both the AGN XLF and and $P_\mathrm{AGN}$ are highly degenerate, being dependent on several input parameters. On the other hand, the AGN bias as a function of stellar mass depends on one single parameter, once $Q$ has been fixed, and in turn also the AGN bias at fixed BH mass depends only on the $M_{\mathrm{bh}}-M_{\mathrm{star}}$ relation, once both $Q$ and the $M_{\mathrm{star}}-M_h$ relation have been fixed.

\begin{figure*}
	\plotone{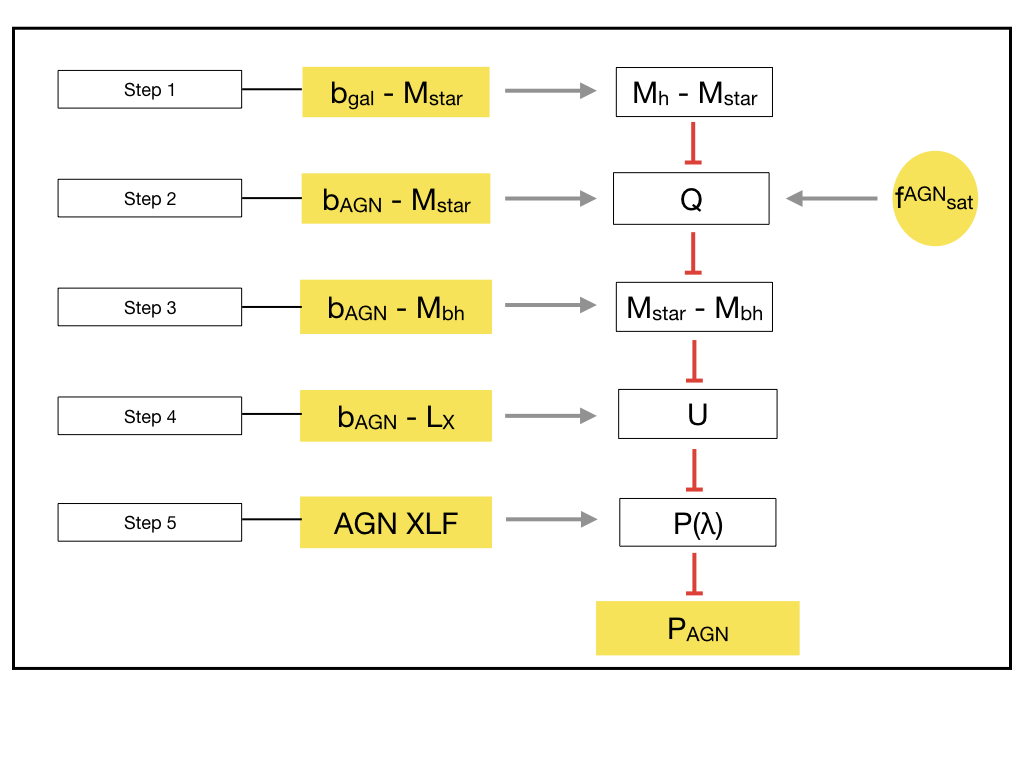}
 	\caption{\footnotesize Sketch of how to build realistic AGN mocks. Full yellow boxes refer to the observables considered in this work. The dependence of each observable on one or few input model parameters (empty boxes) is shown as red lines. From the comparison of observationally derived relations and the AGN mock catalog predictions we can constrain (grey row) the input parameters. Additional observables, such as the fraction of satellite AGN (full yellow circle) can help in breaking the degeneracies among the input model parameters.}\label{fig12}
\end{figure*}

Based on the information contained in Figure~\ref{fig11}, in what follows we provide the different steps to
create a robust and realist mock catalog of AGN\@.
As sketched in Figure~\ref{fig12},
the stellar mass--halo mass relation and the $Q$ parameter can be constrained by combining the large-scale clustering as a function of
stellar mass for both galaxies and AGN (Figure~\ref{fig8}, lower panel), at least in the limit in which AGN hosts are a random subsample of all the galaxies of similar stellar mass.
In particular, our results suggest a model with an input $M_{\mathrm{star}}-M_h$ relation
as described in \citet[][]{grylls19} and $Q \le$ 2 is broadly consistent with available data at $z \le 0.1$.

After having fixed the input stellar mass--halo mass relation and $Q$, the AGN large-scale bias as a function of BH mass can be used to derive the input
$M_{\mathrm{bh}}-M_{\mathrm{star}}$ relation. As already shown in
\citet[][]{shankar20},
we found that a model with a BHSMR-Sha16 (debiased) with $Q \le$ 2
better matches the bias estimates as a function of BH mass (Figure~\ref{fig8}, upper panel).

Observational constraints on the AGN duty cycle
can then be derived from the comparison
of the model predictions with the measured AGN
large-scale bias as a function of AGN luminosity
(Figure~\ref{fig10}). A model with BHSMR-Sha16 (debiased),
$Q \le$ 2 and U-SW10 (decr) is able to reproduce the AGN bias as a function of $L_X$, for both a Gaussian or Schechter $P(\lambda$).

Finally, after having fixed the stellar mass--halo mass relation
\citep[][]{grylls19}, BHSMR-Sha16 (debiased), $Q \le$ 2 and U-SW10 (decr), the combination of the AGN XLF and the specific accretion-rate distribution $P_\mathrm{AGN}$ allow us to derive the free parameters of the input Eddington ratio distribution, independently of the exact shape of the input $P(\lambda)$.

Estimates of the fraction of active satellites in
groups and clusters at the redshift of interest
\citep[e.g.][]{allevato12,leauthaud15}
can further help to independently constrain the $Q$ parameter \citep[e.g.,][]{gatti16}.
Additional observables can be considered, such as the
average L$_X$-SFR/M$_{\mathrm{star}}$ relation, which mostly depends on
$P(\lambda$) and on the M$_{\mathrm{star}}$-M$_{\mathrm{BH}}$ relation (Carraro et al., submitted).

Our current work thus reveals the right observables that
we should focus on to break the degeneracies in the model input parameters and it provides the steps to build a robust and realistic AGN mock consistent with many different observables. At the same time, our framework represents an invaluable tool to shed light on the cosmological evolution of BHs, providing key constraints on the underlying scaling relations between BHs, their galaxies and host dark matter halos, along with information on their accretion rates, frequency (the duty cycle), and environmental dependence (via the $Q$ parameter).

\section{Discussion}
\label{sec:disc}

\subsection{Specific accretion-rate distribution}
In this work we showed how observables depend on the input model parameters (Figure~\ref{fig11}) and how to build step-by-step
robust mock catalogs of AGN that minimize the danger of inner degeneracies and include knowledge of the underlying black hole mass and Eddington ratio distributions (Figure~\ref{fig12}).

The first observable we considered is the specific accretion-rate distribution $P_\mathrm{AGN}$, defined as the convolution of the input AGN duty cycle $U$ and normalized Eddington ratio distribution $P(\lambda$).
The $P_\mathrm{AGN}$ distribution has been intensively studied
in the last decade mostly in X-ray selected AGN samples
\citep[e.g.][]{bongiorno16,aird17,aird18,georgakakis17}
and it has been extensively used as the main key
observable to generate data-driven AGN mock catalogs
\citep[e.g.][]{comparat19,aird20}.

However, when using models uniquely tuned on the
measured $P_\mathrm{AGN}$, we miss information
on individual input parameters, such as the AGN duty cycle $U$, the
Eddington ratio distribution and the BH mass -- stellar mass relation.
We in fact showed in the previous Sections that, for a fixed $M_{\mathrm{bh}}-M_{\mathrm{star}}$ relation, widely different combinations of $U$ and $P(\lambda$) can provide very similar specific accretion-rate distributions and AGN XLFs consistent with the data.
Moreover, any specific accretion-rate distribution
$P_\mathrm{AGN}$ that reproduces the AGN XLF,
does not affect the AGN large-scale bias at a given stellar mass. Thus the
$P_\mathrm{AGN}$ distribution and the AGN XLF are not
suited to constrain the input model parameters when used in isolation. 

On the contrary, in this work we explicitly consider
the AGN duty cycle, $P(\lambda$)
and the BH mass--stellar mass relation as distinct
input model parameters, which we tested against
several independent observables, including the large-scale
bias as a function of stellar/BH mass and
X-ray luminosity. In particular, we found that the comparison of
observationally derived $P_\mathrm{AGN}$ with the
predictions of AGN mock
catalogs are in better agreement with models that
assume an input BH mass--stellar mass relation lower in normalization
(which we labelled throughout as BHSMR-Sha16, debiased)
with respect to what usually inferred in the local Universe from early-type galaxies with dynamically measured BHs (which we labelled throughout as BHSMR-SG16, raw).
Our mock also prefers AGN duty cycle decreasing with BH mass (U-SW10)
consistent with what also derived from continuity equation arguments
\citep[e.g.][]{shankar13}.
The agreement with the data, and in particular with the measured $P_\mathrm{AGN}$ function, further improves when the input Eddington ratio distribution depends on the BH mass $P(\lambda,M_{bh})$, for both a Gaussian and Schechter function, or for example assuming a double power-law distribution \citep{yang19}. 
It is worth noticing that, when considered in isolation, the $P_\mathrm{AGN}$ distribution can be also reproduced, at least at lower luminosities/stellar masses log$\lambda \le $33,
by using in input BHSMR-SG16 (raw) and U-M19 (incr) or U-G10 (const)
(for both a Gaussian or a Schechter $P(\lambda$)). This degeneracy can be broken by testing the model against additional independent observable, most notably the AGN large-scale clustering.

\subsection{Bias vs M$_{star}$/M$_{bh}$}

The second key observable to consider is indeed the AGN large-scale bias as a function of both stellar mass and BH mass, which is not affected by the input AGN duty cycle and $P(\lambda$). 
The large-scale bias as a function of the host galaxy stellar mass 
is set by the stellar mass--halo mass relation and it is independent of the AGN model (i.e. the AGN duty cycle, BH mass--stellar mass relation and Eddington ratio distribution) as long as, as discussed above, the AGN host galaxies are a random subsample of the galaxies with similar stellar mass. \citet{georgakakis19} also found that the level of clustering of AGN samples primarily correlates with the stellar masses of their host galaxies, rather than their instantaneous accretion luminosities.

As shown in
\citet[][]{shankar20},
the AGN large-scale bias as a function of BH mass can instead effectively be used to put constraints on the BH mass--stellar mass relation and the parameter $Q$, the ratio of satellite and central active galaxies/BHs.
In detail,
\citet[][]{shankar20}
found that the observed bias of AGN at $z = 0.25$
\citep[][]{krumpe15}
can be reproduced by assuming BHSMR-Sha16 (debiased) and $Q \le 2$,
which corresponds to satellite AGN fractions f$_{sat}^\mathrm{AGN} \le 0.15$.
A similar value (f$_{sat}^\mathrm{AGN} \sim$ 0.18) has been suggested by
\citet[][]{leauthaud15}
for COSMOS AGN at z $<$ 1.
\citet[][]{allevato12}
performed direct measurement
of the HOD for COSMOS AGN based on the mass function
of galaxy groups hosting AGN and found that the duty
cycle of satellite AGN is comparable or slightly larger
than that of central AGN, i.e. Q $\le$ 2.
A very low value of the Q parameter would be in line with quasars hosted in central galaxies that more frequently undergo mergers with other galaxies
\citep[][]{hopkins08}.
On the other hand,
a relatively high value of Q would suggest that other triggering mechanisms other than mergers, such as secular processes and bar instabilities, are equally, or even more efficient, in producing luminous AGN
\citep[e.g.][]{georgakakis09,allevato11,gatti16}.
The semi-empirical model used in \citet{georgakakis19} 
for populating halos with AGN does not distinguish
between central and satellite active BHs, i.e. effectively their model adopt $Q=1$, which implies a satellite fraction of f$_{sat}^\mathrm{AGN} \sim$ 10-20\%. \citet{georgakakis19} claim, as also found here, that the fair agreement of their mocks with the observationally derived AGN HOD 
(e.g., Allevato et al. 2012, Miyaji et al. 2011, Shen et al. 2013) supports low values of the $Q$ parameter.

We find that a model with an input BHSMR-Sha16 (debiased) and $Q \le 2$ does indeed better match the large-scale bias as a function of BH mass of X-ray AGN at z $<$ 0.1
\citep[][]{powell18},
further extending the results of
\citet[][]{shankar20}
even at lower redshifts. Additionally, the same model is in better agreement with observationally inferred $P_\mathrm{AGN}$ distributions. This model also assumes: (i) a stellar mass -- halo mass relation as derived in
\citet[][]{grylls19},
which reproduces the most recent estimates of the local galaxy stellar mass function by
\citet[][]{bernardi17},
and it is consistent, as shown in Figure~\ref{fig8}, with the large-scale clustering of local central AGN in SDSS 
\citep[][]{zhang20},
and SDSS galaxies with photometry from \citet{dominguez18};
(ii) A parameter $Q \le$ 2 as suggested by observations of the AGN satellite fraction at low redshifts
\citep[e.g.][]{allevato12,leauthaud15}.

A model with BHSMR-SG16 (raw) with U-M19 (incr) or U-G10 (const) would instead require high values of the $Q$ parameter ($Q>3$) and/or an input stellar mass -- halo mass relation as derived by
\citet[][]{moster13}.
More importantly, the latter model is inconsistent with the large-scale bias versus X-ray luminosity
inferred for X-ray AGN at similar redshift
\citep[e.g.][]{krumpe18,powell18},
independently of the choice of the input $P(\lambda$) distribution.

It is worth noticing that our results in terms of AGN large-scale bias as a function of stellar/BH mass are not affected by the choice of a BH mass dependent input $P(\lambda,M_{bh})$. On the contrary, the covariant scatter smooths out the large-scale bias dependence on the stellar mass for mock AGNs, especially at $M_{star} > 10^{11} M_{\odot}$. Currently available bias estimates of SDSS AGN \citep{zhang20} favor models for the creation of mock catalogs without covariant scatter, at least at z $\sim$ 0.1. In the near future, clustering measurements of AGN that extend up to $M_{star} > 10^{11} M_{\odot}$ will allow us to confirm these results (see Viitanen et al. (submitted) for a more comprehensive discussion of the role of covariant scatter at z $\sim$ 1.)

\subsection{Bias vs L$_{X}$}

The large-scale AGN bias as a function of X-ray luminosity represents an additional crucial and powerful diagnostic to constrain viable AGN models, as it is strongly dependent on the input AGN duty cycle, but weakly dependent on the input stellar mass -- halo mass relation or $P(\lambda)$ distribution (Figure 8). 
The large-scale bias as a function of luminosity for mock AGNs 
has been investigated in \citet{georgakakis19} and \citet{aird20} at different redshifts. Their semi-empirical models predict negligible, or extremely weak, dependence of the AGN clustering on
accretion luminosity. We also found an almost constant relation 
between the bias and the AGN X-ray luminosity, especially when using 
U-SW10 (decr), independently of the particular choice of the stellar mass--halo mass relation, BH mass--stellar mass relation, and P($\lambda$).

Measurements of the bias dependence on $L_X$ for X-ray selected AGN at z $\le$ 0.1 \citep[e.g.][]{krumpe18,powell18}
can be reproduced in the models presented in this work assuming
(i) U-SW10 (decr) and BHSMR-Sha16 (debiased)
with $Q \le$ 2;
(ii) or BHSMR-SG16 (raw) with $Q >$ 3. This is valid for both a Schechter
or Gaussian $P(\lambda$) or $P(\lambda,M_{bh}$). However in the latter case (ii), the corresponding
specific accretion-rate distribution $P_\mathrm{AGN}$ would be almost
one order of magnitude higher than observations.
\citet{georgakakis19} and \citet{aird20} also compared AGN bias estimates and/or halo mass as a function  of luminosity with mock AGN predictions. At redshift z $\sim$ 0.25--0.3, they found a small offset with respect to measurements that require revisiting some of their model input assumptions or be due to selection effects of specific
samples, e.g. redshift interval and X-ray flux limits.

As discussed in the previous section, \emph{only} the combination
of all the observables, namely the AGN XLF, the $P_\mathrm{AGN}$
distribution, the AGN large-scale
bias as a function of stellar/BH mass and $L_X$ can break the degeneracy in the
input model parameters and ensure the creation of realistic AGN mock catalogs.

\section{Conclusions}
\label{sec:conc}

In this work we describe a step-by-step methodology to create robust, transparent and physically motivated AGN mock catalogs that can be safely used for extra-galactic large-scale surveys and as a testbed for cosmological models of BH and galaxy co-evolution. Our methodology, summarized in Figure 11, allows to minimise the danger of degeneracies and to pin down the underlying physical properties of BHs in terms of their accretion distributions and links to their host galaxies. More specifically, we find that:

\begin{itemize}
    \item The AGN XLF and the specific accretion-rate distribution $P_\mathrm{AGN}$
    depend on the input $M_{\mathrm{bh}}$-$M_{\mathrm{star}}$ and $M_{\mathrm{star}}$-$M_h$ relations,
    Eddington ratio distribution $P(\lambda$) and AGN duty cycle $U$, and are independent of the particular choice of $Q$, parametrizing the ratio between satellite and central AGN at a given host galaxy stellar mass.

    \item The clustering at fixed stellar mass only depends
    on the $M_{\mathrm{star}}$-$M_h$ relation relation and the $Q$ parameter.

    \item The clustering at fixed BH mass only depends
    on the $M_{\mathrm{bh}}$-$M_{\mathrm{star}}$, the $M_{\mathrm{star}}$-$M_h$ relations and the $Q$ parameter.


    \item All AGN mocks built on empirically-based $M_{\mathrm{star}}$-$M_h$ relations, will broadly match the AGN clustering at a given stellar mass, provided the AGN hosts are a random subsample of the underlying galaxy population of the same stellar mass.
    

    
    \item A large variety of specific accretion-rate distributions $P_\mathrm{AGN}$, defined as the convolution of the normalized Eddington ratio distribution $P(\lambda$) and the AGN duty cycle $U$, can reproduce the AGN XLF even if characterized by widely different underlying $M_{\mathrm{bh}}-M_{\mathrm{star}}$ and/or duty cycles and/or $P(\lambda$).

    \item Only the combination with additional observables, most notably the AGN large-scale bias as a function of BH mass and X-ray luminosity, can break the (strong) degeneracies in the input model parameters.
\end{itemize}

The results listed above indeed imply strong degeneracies among the input parameters used to create mock catalogs of AGN\@.
Having characterised a given $P_\mathrm{AGN}$ that, by design, observationally fits the AGN XLF, does not guarantee a unique and valid solution to create realistic AGN mocks, even when we consider the clustering at fixed stellar mass, simply because the latter is mostly dependent on $Q$ and on the $M_{\mathrm{star}}-M_h$ relation. 


The AGN large-scale bias, as a function of both BH
mass and X-ray luminosity, is a crucial diagnostic for all AGN models.
In particular, a model with an input stellar mass -- halo mass
relation calibrated from detailed abundance matching
\citep[e.g.][]{grylls19},
a $M_{\mathrm{bh}}-M_{\mathrm{star}}$ with lower normalizations than those usually inferred for dynamically measured local BHs \citep[e.g.,][]{reines15,shankar16}, an AGN duty cycle
decreasing with BH mass \citep[e.g.,][]{schulze10}, combined with the assumption that central and satellite BHs of equal mass share similar probabilities of being active (i.e. $Q\le2$), generates a mock
catalog of AGN that matches the observationally constrained AGN XLF,
$P_\mathrm{AGN}$ and AGN large-scale bias as a function of the stellar/BH mass
and X-ray luminosity at z $\le$0.1. We stress that the methodology outlined
in this work is of wide applicability and we expect it to hold at all redshifts, thus allowing to constrain the evolution in the BH scaling relations, duty cycles, and
Eddington ratio distributions (e.g. Viitanen et al., submitted).

Additional observables, not included in the present work, can also be considered to set stronger/additional constraints on the input model parameters, for instance the average L$_X$-SFR/M$_{\mathrm{star}}$ relation, which mostly depends on the input Eddington ratio distribution
$P(\lambda$) and on the M$_{\mathrm{bh}}$-M$_{\mathrm{star}}$ relation (Carraro et al.\ submitted).
Estimates of the fraction of active satellites in groups and clusters at low redshift
\citep[e.g.][]{allevato12,leauthaud15}
are also key observables to independently constrain the $Q$ parameter \citep[e.g.,][]{gatti16}.

Our present study provides a complete framework to build
robust and realistic AGN theoretical samples consistent with diverse and largely independent
observables, and it is capable of setting strong constraints on the main parameters controlling the growth of BHs in galaxies. Our work can thus provide key insights into cosmological galaxy evolution models whilst defining a clear strategy to produce robust galaxy mock catalogues for the imminent large-scale galaxy surveys such as Euclid and LSST\@.

\acknowledgments

FS acknowledges partial support from a Leverhulme Trust Research Fellowship.
AV is grateful to the Vilho, Yrjö and Kalle Väisälä Foundation of the Finnish Academy of Science and Letters. 
C. Marsden acknowledges ESPRC funding for his PhD. 
FS warmly thanks David Weinberg for many useful discussions.
We acknowledge extensive use of the Python libraries Colossus, Corrfunc, astropy, matplotlib, numpy, pandas, and scipy.
The MultiDark Database used in this paper and the web application providing online access to it were constructed as part of the activities of the German Astrophysical Virtual Observatory as result of a collaboration between the Leibniz-Institute for Astrophysics Potsdam (AIP) and the Spanish MultiDark Consolider Project CSD2009-00064. The Bolshoi and MultiDark simulations were run on the NASA’s Pleiades supercomputer at the NASA Ames Research Center. The MultiDark-Planck (MDPL) and the BigMD simulation suite have been performed in the Supermuc supercomputer at LRZ using time granted by PRACE\@.

\section*{Data Availability}

The MultiDark ROCKSTAR halo catalogues are available in the CosmoSim database at \url{https://www.cosmosim.org/}. Other data underlying this article will be shared on reasonable request to the corresponding author.

\bibliographystyle{aasjournal}
\bibliography{SEMs.bib}

\end{document}